\documentclass[pra,twocolumn,amsfonts,amsmath,floatfix]{revtex4}
\usepackage{epsfig,amsmath,graphicx,psfrag}

\newcommand{\lt}{\left(}
\newcommand{\rt}{\right)}
\newcommand{\lqu}{\left[}
\newcommand{\rqu}{\right]}
\newcommand{\lgr}{\left\{}
\newcommand{\rgr}{\right\}}

\newcommand{\be}{\begin{equation}}
\newcommand{\ee}{\end{equation}}
\newcommand{\ba}{\begin{eqnarray}}
\newcommand{\ea}{\end{eqnarray}}
\newcommand{\fr}{\frac}
\newcommand{\nn}{\nonumber}
\newcommand{\se}{\section}
\newcommand{\sse}{\subsection}
\newcommand{\ssse}{\subsubsection}

\begin{document}

\title{Symplectic approach to the amplification process in a nonlinear fiber}

\author{G. Ferrini$^1$}
\email{giulia.ferrini@spectro.jussieu.fr}
\author{I. Fsaifes$^2$}
\author{T. Labidi$^2$}
\author{F. Goldfarb$^2$}
\author{N. Treps$^1$}
\author{F. Bretenaker$^2$}
\affiliation{$^1$ Laboratoire Kastler Brossel, CNRS-Universit\'e Pierre et Marie Curie-Paris 6 - ENS, 4 place Jussieu, 75252 Paris, France\\
$^2$ Laboratoire Aim\'e Cotton, CNRS-Universit\'e Paris Sud 11-ENS Cachan, Campus d'Orsay, 91405 Orsay, France}
\date{\today}

\begin{abstract}

We analyze the amplification processes occurring in a nonlinear fiber, either driven with one or two pumps. After determining the solution for the signal and idler fields resulting from these amplification processes, we analyze the physical transformations that these fields undergo. To this aim, we use a Bloch-Messiah decomposition for the symplectic transformation governing the fields evolution. 
Although conceptually equivalent to other works in this area [McKinstrie and Karlsson, Opt. Expr. \textbf{21}, 1374 (2013)], this analysis is intended to be particularly simple, gathering results spread in the literature, which is useful for guiding practical implementations. Furthermore, we present a study of the correlations of the signal-idler fields at the amplifier output. We show that these fields are correlated, study their correlations as a function of  the pump power, and stress the impact of these correlations on the amplifier noise figure. Finally, we address the effect of losses. We determine whether it is advantageous to consider a link consisting in an amplifying non-linear fiber, followed by a standard fiber based lossy transmission line, or whether the two elements should be reversed, by comparing the respective noise figures.

\end{abstract}

\maketitle

\section{Introduction}
\label{se:intro}
There is a growing interest for parametric amplification of optically carried signals using gain provided by four-wave mixing in third-order nonlinear fibers~\cite{Marhic_08,Tong_12,Tong_AdvPhot_13}. One of the reasons of this strong activity is that such amplifiers can be operated in the so-called phase sensitive regime, leading in principle to the possibility to amplify the signal without degrading the signal-to-noise ratio~\cite{Caves_82}. Using recently available highly nonlinear fibers (HNLFs) which exhibit low losses~\cite{Agrawal_13}, this has led to recent demonstrations of phase and amplitude regenerations~\cite{Slavik_10} and amplification with signal-to-noise degradation below the 3~dB limit of phase insensitive amplifiers~\cite{Tong_11}.

Different types of theoretical descriptions of phase sensitive amplification have been published. Some of them rely on a classical description of the electromagnetic field~\cite{McKinstrie_04a,Vasilyev_05,Tong_10,Lundstrom_12}. Some others directly calculate the variances of the quantum fields at the output of the amplifier~\cite{McKinstrie_04b,McKinstrie_05,McKinstrie_06,McKinstrie_10, Marhic_12, Marhic_13}. Finally, parametric processes have recently been analyzed using a singular value (Schmidt) decomposition of the transformation matrix~\cite{McKinstrie_09,McKinstrie_13a,McKinstrie_13b}. Important aspects for the design of efficient phase sensitive amplifiers have been discussed, such as the gain~\cite{McKinstrie_04a}, the noise figure~\cite{McKinstrie_04b,Vasilyev_05,McKinstrie_05,Tong_10, Marhic_12, Marhic_13}, the choice between various pumping schemes~\cite{Vasilyev_05,Marhic_13}, the influence of the state of polarization of the different fields~\cite{McKinstrie_06,Marhic_11,McKinstrie_13b}, the influence of gain saturation~\cite{Lundstrom_12}, or the correlations between fields~\cite{McKinstrie_10}, etc.

Our aim here is to give a conceptually and mathematically simple description of most of these various aspects of parametric amplification. Our approach is based on a Bloch-Messiah decomposition~\cite{Caves_85,Kolobov_99,Braunstein_05} of the symplectic transformation governing the evolution of the fields. We show that this approach permits to emphasize the physical interpretation of these transformations. This allows us not only to reinterpret the predictions of other approaches, but it can also be useful in practical situations, for example to predict the state of the field at the output of the amplifier in terms of the mean-fields, the fields fluctuations and the correlations between signal and idler fields when these two modes are present at the output.

\begin{figure}[h]
  \centerline{\includegraphics[width=1.0\columnwidth]{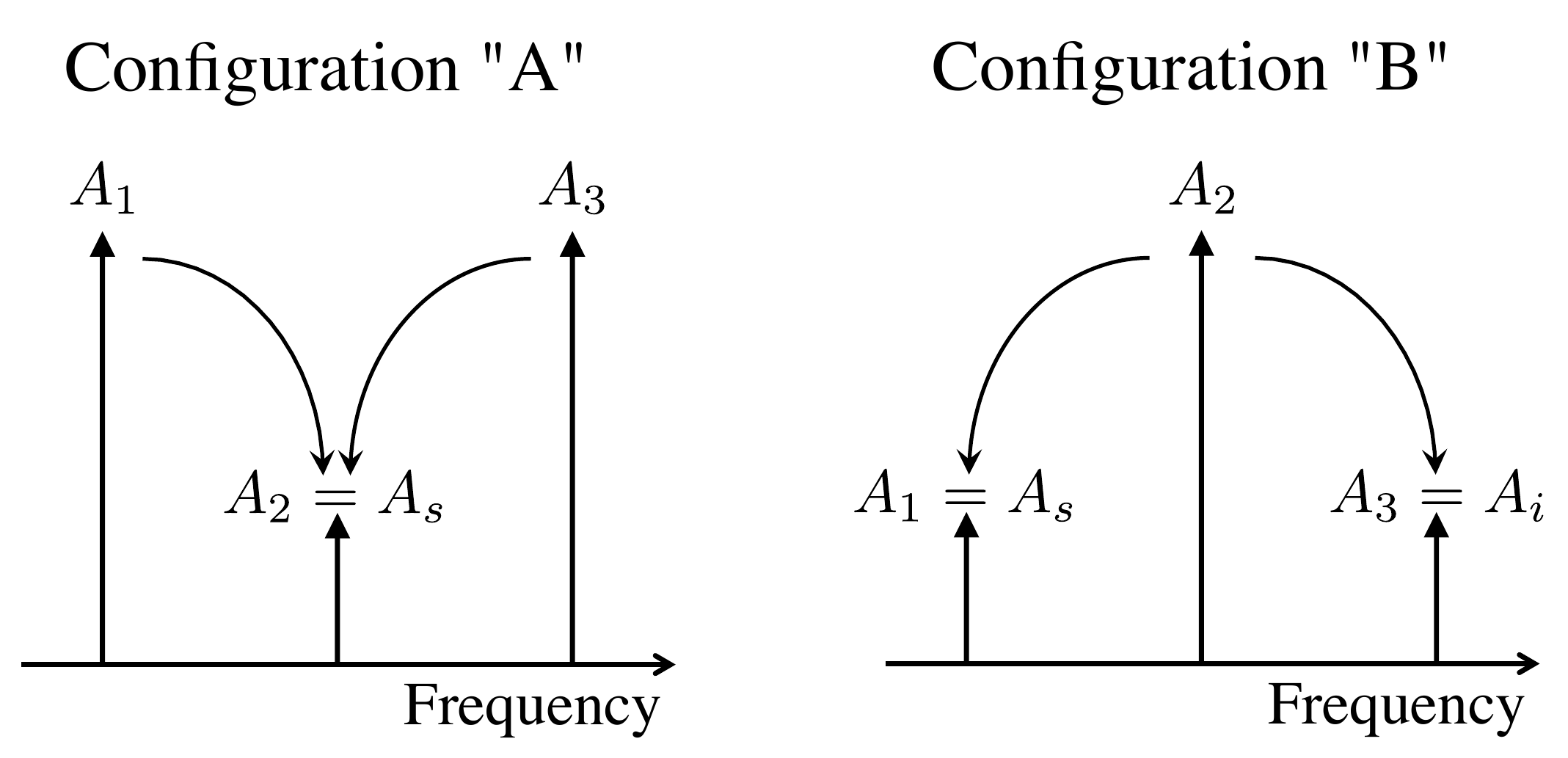}}
  \caption{Schematic representation of the two considered configurations for parametric amplification based on four-wave mixing. In configuration ``A'', $A_2=A_s$ is the amplitude of the degenerate signal and idler while $A_1$ and $A_3$ are the amplitudes of the two pumps. Conversely, in configuration ``B'', $A_2$ holds for the two degenerate pumps while $A_1=A_s$ and $A_3=A_i$ are the amplitudes of the signal and idler, respectively. }\label{Fig01}
 \end{figure}

In the following, we consider the two usual architectures (see Fig.\;\ref{Fig01}) in which either the signal and the idler or the two pumps are frequency degenerate.

The paper is organized as follows. In Section II we set the problem and remind the derivation of the classical expressions for the parametric gains. This allows us to determine the matrices  describing the amplifier in the two configurations. Section III gives the Bloch-Messiah decomposition of the transfer matrices with illustrations in the different configurations. Section IV derives the optimum input condition. Some applications to the derivation of the output signal-to-noise ratio are given in Section V. A particular emphasis is put on the influence of the correlations between the various fields on the analysis of the noise figure of the amplifier. Finally, section VI is devoted to loss management in such phase sensitive amplifiers.

\section{Classical parametric gains: matrix approach} \label{eqs_classiche}
The aim of this Section is to derive the symplectic transfer matrix for the four-wave mixing process in the two considered configurations. We perform this derivation in the framework of the classical approximation. Quantum fields will be considered in the following Sections. The linearity of the system will enable the quantum fields to be equivalently described by the classical solution~\cite{Loudon_00, GAF}.

In general, four-wave mixing in fibers involves four different frequencies. However, the simplest configurations occur when two of these frequencies are equal, leading to the interaction of three different fields. We will note the complex amplitudes of these classical fields $A_i$, their frequencies $\omega_i$, and their wave vectors $\beta_i$, with $i=1,2,3$.
 
The evolution of the complex amplitudes is governed by the following set of differential equations~\cite{McKinstrie_04a, Chen_12, Hansryd_02, Lundstrom_12}:
\ba
\label{eq:4-wave-mixing}
\fr{\mathrm{d} A_1}{\mathrm{d}z} \hspace{-0.1cm} &=& \mathrm{i} \gamma \lqu (|A_1|^2 + 2|A_2|^2 + 2 |A_3|^2) A_1\right.\nn\\
&&\left. + A_2^2 A_3^*\mathrm{e}^{\mathrm{i} \Delta\beta z} \rqu ,     \\
\fr{\mathrm{d} A_2}{\mathrm{d}z} \hspace{-0.1cm} &=& \mathrm{i} \gamma \lqu (2|A_1|^2 + |A_2|^2 + 2 |A_3|^2) A_2 \right.\nn\\
&&\left.  + 2 A_1 A_3 A_2^* \mathrm{e}^{-\mathrm{i} \Delta\beta z} \rqu ,  \\
\fr{\mathrm{d} A_3}{\mathrm{d}z} \hspace{-0.1cm} &=& \mathrm{i} \gamma \lqu (2|A_1|^2 + 2|A_2|^2 +  |A_3|^2) A_3\right.\nn\\
&&\left.  + A_1^* A_2^2  \mathrm{e}^{\mathrm{i} \Delta\beta z}  \rqu ,
\ea
where $\Delta\beta= 2 \beta_2 - \beta_3 - \beta_1$ is the linear phase mismatch and $\gamma$ the fiber nonlinear coefficient. 

The two configurations to which we apply these equations are schematized in Fig.\,\ref{Fig01}. For the configuration labeled ``A'', in which there are two non-degenerate pumps (while signal and idler are degenerate), we interpret the fields $A_1$ and $A_3$ as the strong pumps, while $A_2\equiv A_s$ represents the degenerate signal and idler fields. For the configuration labeled ``B'', the two pumps are degenerate with amplitude $A_2$ and $A_1\equiv A_s$ and $A_3 \equiv A_i$ correspond to the signal and idler, respectively.

Following Ref.~\cite{Agrawal_13}, we are going to provide a solution for the evolution of the complex amplitudes of these fields in their respective two configurations.

\subsection{Configuration ``A'': Degenerate Signal and Idler}

Consider first the case of two non-degenerate pumps.
In the undepleted pump approximation $|A_{1,3}|^2 \equiv P_{1,3}$ are constant, and with the identification $A_2(z) = A_s(z)$, where $z$ is the propagation distance along the fiber, we derive the following solution (see Appendix \ref{app:solA}):
\ba
&A_s = \mu A_{s0} + \nu  A_{s0}^* ,  \label{eq:solB1} \\
&\mu =  \lt \cosh{gz} +   \mathrm{i}\fr{\kappa}{2g} \sinh{gz}  \rt \mathrm{e}^{\mathrm{i} \delta z} , \label{eq:solB2} \\
&\nu =  \fr{2 \mathrm{i} \gamma}{g} A_{1}(0) A_{3}(0)   \sinh{gz} \, \,   \mathrm{e}^{\mathrm{i} \delta z}, \label{eq:solB3}
\ea
where for the sake of conciseness we note here and in the rest of the paper $A_s \equiv A_{s}(z) $ (the dependence on the spatial coordinate $z$ is implicit in the coefficients $\mu$ and $\nu$), $A_{s0} \equiv A_s(0)$, and where
\ba
\kappa =  \Delta\beta + \gamma (P_{1} + P_{3})
\ea
 is the nonlinear phase mismatch, 
 \ba
 g = \sqrt{4 \gamma^2 P_{1} P_{3} - (\kappa/2)^2 }
 \ea
 is the parametric gain coefficient, and where
 \ba
 \delta = \fr{  3 \gamma (P_{1} + P_{3}) - \Delta\beta }{2} 
 \ea
 is the effective phase mismatch. 
 Eqs. (\ref{eq:solB1}-\ref{eq:solB3}) can be recast into the matrix form
\be
\label{eq:matrici}
\left( 
\begin{array}{cccccccc}
A_{s}  \\
A^*_{s} \\
\end{array} 
\right) =
\left( 
\begin{array}{cccccccc}
\mu & \nu \\
\nu^* & \mu^* \\
\end{array} 
\right)
\left( 
\begin{array}{cccccccc}
A_{s0}  \\
A^*_{s0} \\
\end{array} 
\right).
\ee
It is readily verified that $|\mu|^2 - |\nu|^2 = 1$, which reflects the fact that the matrix in Eq.\,(\ref{eq:matrici}) belongs to the symplectic group~\cite{Dutta, Simon_88}. Indeed, for the moment, we did not take losses into account, and the linearity of the system translates into the input-output relations Eq.\,(\ref{eq:matrici}). 
At the quantum level, the property $|\mu|^2 - |\nu|^2 = 1$ ensures that Eq.\,(\ref{eq:solB1}) holds with no need to add extra noise terms coming from the amplification process [see Eq.\,(3.15) in Ref.~\cite{Caves_82}]. The evolution described in Eq.\,(\ref{eq:solB1}) denotes hence a ``noiseless amplifier".

From Eq.(\ref{eq:solB1}) the usual power gain can be computed  and results in 
\begin{widetext}
\ba
\label{eq:clax_gain}
G &=& \fr{|A_s|^2}{|A_{s0}|^2} = \fr{|\mu A_{s0} + \nu A^*_{s0}|^2}{P_{s0}} = |\mu|^2 +  |\nu|^2 + 2 |\mu| |\nu| \cos (\theta_\mu - \theta_\nu - 2 \theta_{s0})   \nn \\ 
&=& 1 + \lqu 1 + \fr{\kappa^2 + 16 \gamma^2 P_1 P_3 + 8 \kappa \gamma \sqrt{P_1 P_3} \cos \xi_{\text{rel}}}{4 g^2} \rqu \sinh^2 gz  +  \fr{2 \gamma}{g} \sqrt{P_1 P_3} \sin \xi_{\text{rel}} \sinh 2 gz,
\ea
\end{widetext}
where the last expression is obtained after few algebraic steps, where we have defined $|A_{s0}|^2 = P_{s0} = P_{20}$ and $\mu = |\mu | e^{i \theta_\mu}$, $\nu = |\nu | e^{i \theta_\nu}$ and where 
\ba
\xi_{\text{rel}} = 
2 \theta_{s0} - \theta_{10} - \theta_{30},
\ea
is the relative phase between the waves. The $\theta_{0i}$'s ($i=1,2,3$) here are the phases of the three input fields:
\ba
A_{0i}=|A_{0i}|\mathrm{e}^{\mathrm{i}\theta_{0i}},
\ea 
and $\theta_{s0}=\theta_{20}$ is the phase of the input signal field. Note that one can choose the phase reference by setting the phase of one of the two pumps to zero, e.g., $\theta_{10} = 0$.
From Eq. (\ref{eq:clax_gain}) we see that the maximal and minimal gains are respectively 
\ba
\label{eq:clax_gain_maxmin}
G_{\text{max}} &=& ( |\mu| +  |\nu|)^2, \mbox{            for         }  \theta_\mu - \theta_\nu - 2 \theta_{s0} = 2 k \pi\ ,   \\
G_{\text{min}} &=& ( |\mu| -  |\nu|)^2, \mbox{            for         }  \theta_\mu - \theta_\nu - 2 \theta_{s0} = (2k'+1) \pi \nn, \\
\ea
with $k,k'$ integers, as found in Ref. \cite{Tong_11}.

Defining the quadratures $X_{s}$ and $Y_{s}$ as
\ba
\label{def_quadratures1}
A_{s} &\equiv& X_{s} + \mathrm{i}Y_{s},\\
  \hspace{0.5 cm}  A_{s0}& \equiv& X_{s0} + \mathrm{i} Y_{s0}
\ea
allows to rewrite Eq.\,(\ref{eq:matrici}) as 
\ba
\label{eq:simpl_quadr1}
\left( 
\begin{array}{cccccccc}
X_{s}  \\
Y_{s} \\
\end{array} 
\right) &=& 
\left( 
\begin{array}{cccccccc}
\mathcal{R}e \lqu \mu + \nu \rqu & - \mathcal{I}m[\mu - \nu] \\
\mathcal{I}m[\mu + \nu] & \mathcal{R}e[\mu - \nu]  \\
\end{array} 
\right)
\left( 
\begin{array}{cccccccc}
X_{s0}  \\
Y_{s0} \\
\end{array} 
\right) \nn \\
&\equiv&
M
\left( 
\begin{array}{cccccccc}
X_{s0}  \\
Y_{s0} \\
\end{array} 
\right).
\ea
In Section III we will precisely analyze the effect of the transformation expressed by Eq.\,(\ref{eq:simpl_quadr1}) on the input fields.

\subsection{Configuration ``B'': Degenerate Pumps}

Consider now the case of two degenerate pumps, giving rise to a signal and an idler fields, as depicted as configuration ``B'' in Fig.\ \ref{Fig01}.
Using an approach analogous to the one used for configuration ``A'', we provide a solution of Eq.(\ref{eq:4-wave-mixing}) in the undepleted pump approximation, where $A_1(z) = A_s(z)$ and $A_3(z) = A_i(z)$. This reads (see Appendix \ref{app:solB} for the details of the calculation):
\ba
 &A_s = \mu A_{s0} + \nu  A_{i0}^* ,   \label{solA1}\\
& \mu =  \lt \cosh{gz}  +   \mathrm{i}\fr{\kappa}{2g} \sinh{gz}  \rt  \mathrm{e}^{\fr{ \mathrm{i} \beta z}{2}} , \label{solA2} \\
& \nu =  \fr{ \mathrm{i} \gamma  }{g}A_2^2 (0) \sinh{gz}\   \mathrm{e}^{\fr{ \mathrm{i}  \beta z}{2}},\label{solA3}
\ea
where
\be
\kappa =  2 \gamma P_2 - \Delta\beta 
\ee
is the nonlinear phase mismatch,
\be
g = \sqrt{ \gamma^2 P_2^2 - \lt \fr{\kappa}{2} \rt^2 }
\ee
is the parametric gain coefficient and where newly $A_s$ stands for $A_s(z)$.
Again, it is readily verified that $|\mu|^2 - |\nu|^2 = 1$.
An analogous solution can be derived for the idler mode, yielding
\be
\label{solAbis}
 A_i = \nu A_{s0}^* + \mu  A_{i0}.
\ee

By comparing Eq.\,(\ref{eq:solB1}) with Eq.\,(\ref{solA1}), one can see that the coefficients $\mu$ and $\nu$  in the two configurations are related by the following mapping:
\ba
& 2 \sqrt{P_1 P_3} \rightarrow P_2, \label{eq:mapping1} \\
& \theta_{10} + \theta_{30} \rightarrow 2 \theta_{20}, \label{eq:mapping2}\\
& \delta \rightarrow \fr{\Delta\beta}{2}.\label{eq:mapping3}
\ea

Analogously to Eq.\,(\ref{eq:clax_gain}) for configuration ``A'', the power gain in the configuration ``B'' can be computed. For the moment, we give the expression of the power  gain with respect to the signal only. This choice will be motivated in Sec.\ref{se:snrB} where we compute the full noise figure considering both signal and idler fields. 
The signal power gain thus results in 
\begin{widetext}
\ba
\label{eq:gps2}
G &=& \fr{|A_s|^2}{|A_{s0}|^2} = \fr{|\mu A_{s0} + \nu A^*_{i0}|^2}{P_{s0}} = |\mu|^2 +  \eta^2 |\nu|^2 + 2 \eta |\mu|  |\nu| \cos (\theta_\mu - \theta_\nu + \theta_{s0} + \theta_{i0} ) \nn \\ 
&=&    1 + \lqu 1 + \fr{\kappa^2 + 4 \gamma^2 P_2^2 \eta^2 +  4 \kappa \gamma \eta P_2 \cos \xi_{\text{rel}}}{4 g^2} \rqu \sinh^2 gz  +  \fr{ \gamma}{g} \eta P_2 \sin \xi_{\text{rel}} \sinh 2 gz ,
\ea
\end{widetext}
 where we have defined $|A_{s0}|^2 = P_{s0} $ and $|A_{i0}|^2 = P_{i0} $, $\mu = |\mu | e^{i \theta_\mu}$, $\nu = |\nu | e^{i \theta_\nu}$,
 \be
 \xi_{\text{rel}} =  \theta_{s0} + \theta_{i0} - 2 \theta_{20},
 \ee
and
\be
\eta = \sqrt{ \fr{P_{i0}}{P_{s0}}} .
\ee
From Eq. (\ref{eq:gps2}) we see that we obtain the maximal and minimal gains for $\eta = 1$:
\ba
\label{eq:gps2_maxmin}
G_{\text{max}} &=&( |\mu| +  |\nu|)^2, \mbox{         for         }  \theta_\mu - \theta_\nu + \theta_{s0} + \theta_{i0}  = 2 k \pi\ , \\
G_{\text{min}} &=& ( |\mu| -  |\nu|)^2, \mbox{         for         }  \theta_\mu - \theta_\nu + \theta_{s0} + \theta_{i0}  = (2 k'+1) \pi ,\nn\\
\ea
with $k,k'$ integers, consistently with what is found in Ref.~\cite{Tong_11}. It is easy to show that the same expression as the first line in Eq. (\ref{eq:gps2}) holds for the idler power gain, provided one replaces $\eta$ by $1/\eta$.

Contrarily to Eq.\,(\ref{eq:solB1}), Eq.\,(\ref{solAbis}) is not in the form of a standard amplification equation in the sense of Ref.~\cite{Caves_82}. However, it can be recast in such a form by introducing the change of basis et each $z$
\ba
A_+(z)& =& \fr{A_s(z) + A_i(z)}{\sqrt{2}} , \label{eq:changing1}\\
A_-(z) &=& \fr{A_s(z) - A_i(z)}{\sqrt{2}}\label{eq:changing2}.
\ea
Upon substitution of Eqs. (\ref{eq:changing1}-\ref{eq:changing2}) in Eq. (\ref{solA1}) one readily obtains
\be
\label{eqboh}
A_{\pm } = \mu A_{\pm 0} \pm \nu A^*_{\pm 0}\ .
\ee
These equations constitute a couple of independent parametric amplifier equations for the two fields $A_+$ and $A_-$~\cite{Caves_82}. Note furthermore that both equations (\ref{eqboh}) are a copy of Eq. (\ref{eq:solB1}), with $\nu \rightarrow - \nu$ for the equation governing the field $A_-$. This second mapping will allow further simplifications in the calculations that we will carry out for configuration ``B''.

As precedently, the quadratures
\be
\label{eq:def_quadr}
A_{\pm}(z) \equiv X_{\pm}(z) + i Y_{\pm}(z)
 \ee
 can be defined, leading to the evolution equation
\ba
\label{eq:simpl_quadr}
\left(  
\begin{array}{cccccccc}
X_{\pm}  \\
Y_{\pm} \\
\end{array} 
\right) &=& 
\left( 
\begin{array}{cccccccc}
\mathcal{R}e \lqu \mu \pm \nu \rqu & - \mathcal{I}m[\mu \mp \nu] \\
\mathcal{I}m[\mu \pm \nu] & \mathcal{R}e[\mu \mp \nu]  \\
\end{array} 
\right)
\left( 
\begin{array}{cccccccc}
X_{\pm 0}  \\
Y_{\pm 0} \\
\end{array} 
\right) \nn \\
& & \equiv
M_\pm
\left( 
\begin{array}{cccccccc}
X_{\pm 0}  \\
Y_{\pm 0} \\
\end{array} 
\right).
\ea

\section{Bloch-Messiah decomposition}

It is very instructive to decompose the matrix expressing the action of the amplifier on the signal and idler modes into a series of fundamental operations, namely a first rotation in the phase space, a squeezing/dilatation and a second rotation~\cite{Dutta, Braunstein_05}. This decomposition is known under the name of Bloch-Messiah (or Euler) reduction, which is the specialization of the singular value decomposition to the case of symplectic matrices.

In the following we will apply the Bloch-Messiah decomposition to the matrices of Eqs. (\ref{eq:simpl_quadr1}) and (\ref{eq:simpl_quadr}) summarizing the effect of the amplification in the ``A" and ``B" configurations, respectively.
A similar approach has been carried out in Ref.~\cite{McKinstrie_13a}, where the Schmidt decomposition of the amplification matrices (\ref{eq:simpl_quadr1}) and (\ref{eq:simpl_quadr}) has been derived. This decomposition is physically equivalent to the Bloch-Messiah one: the precise mathematical link lies in the fact that the columns of the rotation matrices in the Bloch-Messiah decomposition are the Schmidt vectors in the Schmidt decomposition, while the squeezing values correspond to the Schmidt coefficients. Though embedding the same physical meaning, the accent here is placed on the transformations undergone by the modes, which can be visualized by means of the phasor representation. Rotations in the phasor representations for a $\chi^{(2)}$ medium have been studied in Ref. \cite{Kolobov_99}.

Furthermore, our approach is particularly illustrative, as it allows to understand the optimum relative input phase between signal and idler, leading to optimum amplification, in terms of rotations in the phase space. 
We proceed below with the explicit derivation of the decomposition.

\subsection{Configuration ``A''} \label{se:bloch-messiah-configA} 

An instructive way to derive the Bloch-Messiah decomposition consists in wondering for which fields Eq.\,(\ref{eq:simpl_quadr1}) can be recast in the normal form of Ref.~\cite{Caves_82}, i.e. on which fields the amplifier acts with real and positive coefficients $\mu$ and $\nu$. Equation \,(\ref{eq:solB1}) can be rewritten as
\ba
\label{stdeqtris}
A_s = |\mu | \mathrm{e}^{\mathrm{i} \theta_\mu} A_{s0} + |\nu| \mathrm{e}^{\mathrm{i} \theta_\nu} A^*_{s0}.
\ea
Multiplying both members of Eq.\,(\ref{stdeqtris}) by $\exp{\left[-i \lt \fr{\theta_\mu+ \theta_\nu}{2}\rt\right]}$, we immediately obtain
\be
\label{eq:trasf}
A^{''}_{s} = |\mu| A'_{s0} + |\nu|  A^{'*}_{s0}\ ,
\ee
with
\ba
A^{'}_{s0} &\equiv& \mathrm{e}^{\mathrm{i} \lt \fr{\theta_\mu - \theta_\nu}{2} \rt}  A_{s0}\ , \label{eq:bmdem1} \\
A^{''}_{s}  &\equiv& \mathrm{e}^{-\mathrm{i} \lt \fr{\theta_\mu+ \theta_\nu}{2}\rt} A_s\ . \label{eq:bmdem2}
\ea
The same transformation can be expressed in the quadrature representation, yielding
\ba
X^{''}_{s} &=&  (|\mu| +  |\nu|) X'_{s0}\ ,\label{eq:amp_diago1} \\
Y^{''}_{s} &=&  (|\mu| -  |\nu|) Y'_{s0}\ .\label{eq:amp_diago2}
\ea
The introduction of the rotations of Eqs.\ (\ref{eq:bmdem1}, \ref{eq:bmdem2}) has allowed us to identify on which fields the amplifier acts as a genuine squeezer, in the sense of Ref.~\cite{Caves_82}. More explicitly, Eqs.\ (\ref{eq:amp_diago1}, \ref{eq:amp_diago2}) imply that  the input field $A_{s0}$ undergoes the following three successive transformations to become the output field $A_{s}$:
\begin{itemize}
\item A first rotation expressed by Eq.\ (\ref{eq:bmdem1}),
\be
\left( 
\begin{array}{cccccccc}
X'_{s0}   \\
Y'_{s0}  \\
\end{array} 
\right) =
\left( 
\begin{array}{cccccccc}
\cos \lt \fr{\theta_\mu - \theta_\nu}{2} \rt & - \sin \lt \fr{\theta_\mu - \theta_\nu}{2} \rt  \\
\sin \lt \fr{\theta_\mu - \theta_\nu}{2} \rt &  \cos \lt \fr{\theta_\mu - \theta_\nu}{2} \rt \\ 
\end{array} 
\right)
\left( 
\begin{array}{cccccccc}
X_{s0}   \\
Y_{s0}  \\
\end{array} 
\right)\ ;
\ee
\item The amplification with real and positive coefficients expressed by Eqs. (\ref{eq:amp_diago1}, \ref{eq:amp_diago2}):
\be
\left( 
\begin{array}{cccccccc}
X^{''}_{s}   \\
Y^{''}_{s}  \\
\end{array} 
\right) =
\left( 
\begin{array}{cccccccc}
|\mu| +  |\nu| & 0 \\
0 &   |\mu| - |\nu| \\ 
\end{array} 
\right)
\left( 
\begin{array}{cccccccc}
X'_{s0}   \\
Y'_{s0}  \\
\end{array} 
\right)\ ;
\ee
\item A second rotation expressed by Eq.\ (\ref{eq:bmdem2}):
\be
\left( 
\begin{array}{cccccccc}
X_{s}   \\
Y_{s}  \\
\end{array} 
\right) =
\left( 
\begin{array}{cccccccc}
\cos \lt \fr{\theta_\mu + \theta_\nu}{2} \rt & - \sin \lt \fr{\theta_\mu + \theta_\nu}{2} \rt  \\
\sin \lt \fr{\theta_\mu + \theta_\nu}{2} \rt &  \cos \lt \fr{\theta_\mu + \theta_\nu}{2} \rt \\ 
\end{array} 
\right)
\left( 
\begin{array}{cccccccc}
X^{''}_{s}   \\
Y^{''}_{s}  \\
\end{array} 
\right).
\ee
\end{itemize}
To summarize, the symplectic matrix $M$ in Eq.\ (\ref{eq:simpl_quadr1}) can be decomposed as
\ba
\label{eq:b-mvra}
M  &  = & C U  \Sigma  W^T \nn \\
 &   = &C \left( 
\begin{array}{cccccccc}
\cos \theta & - \sin \theta \\
\sin \theta & \cos \theta   \\
\end{array} 
\right)
\hspace{-0.1cm} \left( 
\begin{array}{cccccccc}
|\mu| +  |\nu|& 0 \\
0 & |\mu| -  |\nu|  \\
\end{array} 
\right) \nn \\
&&\times \hspace{-0.1cm} \left( 
\begin{array}{cccccccc}
\cos \phi &  \sin \phi \\
-\sin \phi & \cos \phi   \\
\end{array} 
\right) \ ,
\ea
where $\theta$ and $\phi$ result from Eqs.(\ref{eq:bmdem1}) and (\ref{eq:bmdem2}) in
\ba
\theta& =& \fr{\theta_\mu + \theta_\nu}{2}\ ,\label{eq:sol_angles3} \\
 \phi &=& - \fr{\theta_\mu - \theta_\nu}{2}\ .\label{eq:sol_angles4}
\ea
In the decomposition of Eq. (\ref{eq:b-mvra}), the matrix $C = \pm \mathcal{I}$ is a  two-by-two correction matrix ($\mathcal{I}$ being the identity matrix).
Formally, one has $C = M W \Sigma^{-1}  U^T $ (i.e., whether a sign correction is needed or not can always be determined after that the decomposition has been obtained by directly comparing the sign of $M$ with the one of $U \Sigma W^T$). 

The diagonal matrix $\Sigma$ represents amplification of the projection of the signal amplitude on the direction $X^{\phi}_s= \cos \phi X_{s0} +  \sin \phi Y_{s0}$ with gain $G_{\text{max}} = (|\mu| +  |\nu|)^2 >1$ and  de-amplification of the orthogonal quadrature with gain $G_{\text{min}} =  (|\mu| -  |\nu|)^2<1$, where $G_{\text{max}}$ and $G_{\text{min}}$ are respectively the maximal and minimal amplitude gain as defined in Eq.(\ref{eq:clax_gain_maxmin}).
The product of the two gains is $G_{\text{max}} G_{\text{min}} =  ( |\mu|^2 -  |\nu|^2 )^2 = 1$ yielding noiseless amplification, consistently with the discussion in Ref.~\cite{Caves_82} [see Eq.(1.3) therein]. 

We notice that the second rotation of the Bloch-Messiah decomposition is not experimentally easy to isolate since the definition of the output quadrature
is arbitrary in the absence of an external phase reference.

\begin{figure} [h!]
\centering
\includegraphics[width=1.0\columnwidth]{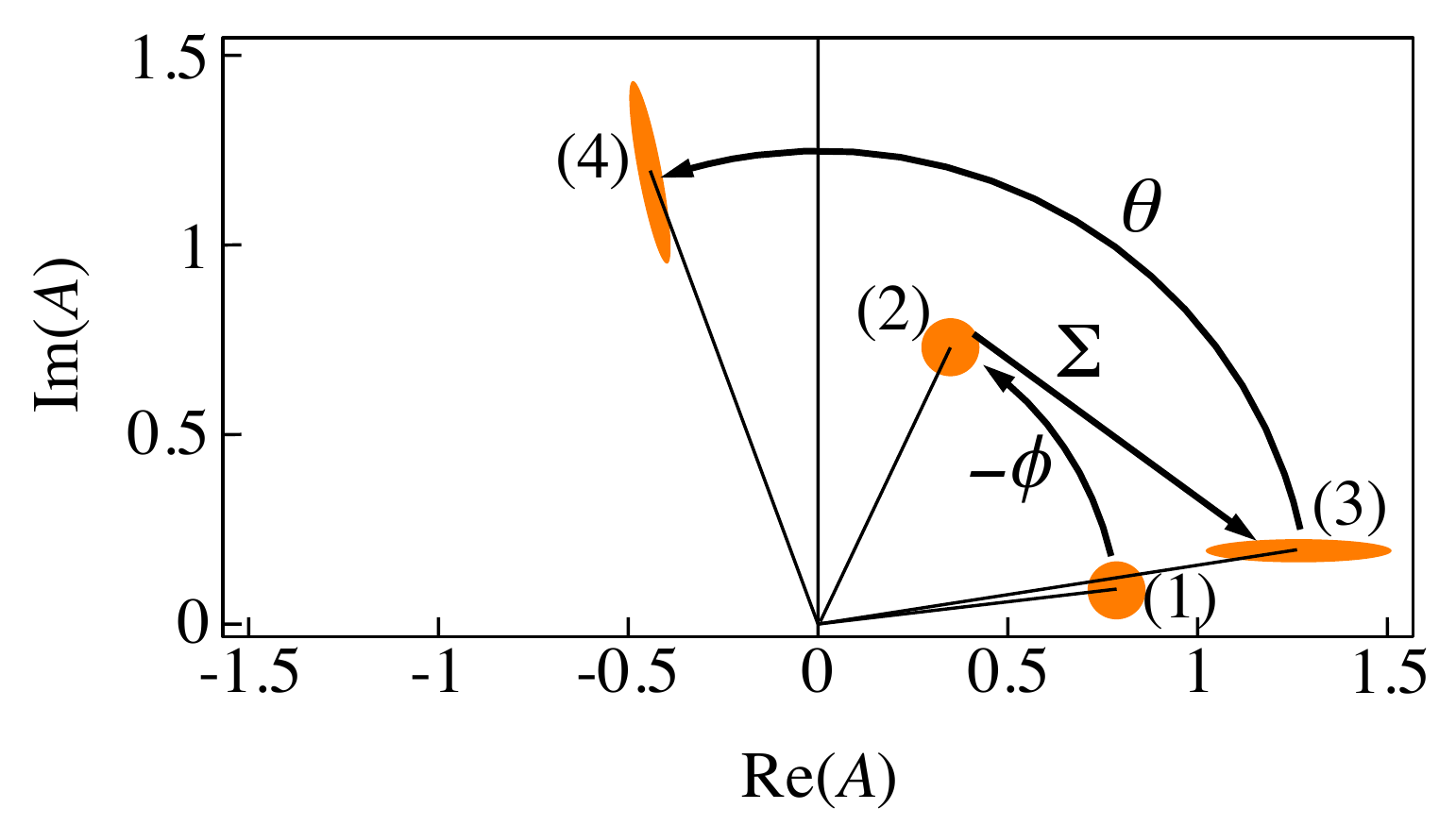}
\caption{Evolution of the field in the phasor representation along the three steps of the Bloch-Messiah decomposition of the parametric amplification in configuration ``A''. The units for the quadratures are arbitrary. The incident field (1), which is in a coherent state $|\alpha_{s0}\rangle$ with $\alpha_{s0}=0.8+i0.1$, undergoes a first rotation by an angle $-\phi$, resulting in field (2). This field is transformed into field (3) by the squeezing operator $\Sigma$, before being rotated by an angle $\theta$ to give the output field (4). The calculations were performed for a fiber length $z=300\ \mathrm{m}$, $\gamma=11.3\times 10^{-3}\,\mathrm{W}^{-1} \mathrm{m}^{-1}$, $\Delta\beta= 4.53\times10^{-11}\,\mathrm{m}^{-1}$, $P_1 =P_3 = 200\,\mathrm{mW}$.}
\label{Fig02}
\end{figure}
The three steps of the Bloch-Messiah decomposition are visualized in Fig.\ \ref{Fig02}, which has been plotted with parameters corresponding to commercially available highly nonlinear fiber (OFS standard highly nonlinear optical fiber~\cite{OFS}). In particular, we take a dispersion slope equal to $0.017\,\mathrm{ps}/(\mathrm{nm}^2 . \mathrm{km})$. In this calculation we suppose that the signal wavelength is located at the zero dispersion wavelength of the fiber (1547 nm in this case) and that the two pumps are located at $\pm0.32\,\mathrm{nm}$ from the pump. Eqs.\ (\ref{eq:sol_angles3}) and (\ref{eq:sol_angles4}) lead to angles $\phi = 57.7\,^{\circ}$ and $\theta = 101.7\,^{\circ}$, and to a gain coefficient $G_{\text{max}}=13.3$. The way these parameters can be directly derived from the experimental data is developed in Appendix \ref{AppendixD}.

\subsection{Configuration ``B''}


The same procedure can be applied to the couple of relations summarized by Eq.\,(\ref{eqboh}), and we have now to define the rotations in input and output for the two fields $A_+$ and $A_-$. The decomposition of Eq.\,(\ref{eq:b-mvra}) exactly holds for the mode $+$, as can be seen by comparing Eq.(\ref{eq:solB1}) with Eq.\,(\ref{eqboh}), and  
we thus obtain 
\be
\label{eq:trasf2p}
A^{''}_{+} = |\mu| A'_{+0} + |\nu|  A^{'*}_{+0}\ ,
\ee
where we have defined
\ba
\label{eq:bmdem2p}
A^{'}_{+0} &\equiv& e^{i \lt \fr{\theta_\mu - \theta_\nu}{2} \rt}  A_{+0}\ , \\
A^{''}_{+}  &\equiv& e^{-i \lt \fr{\theta_\mu+ \theta_\nu}{2}\rt} A_{+}\ .
\ea
For the mode $A_{-}$, in order to define the rotations in the same way as for the mode $+$, we obtain from Eq.\,(\ref{eqboh}) a minus sign in front of $|\nu| $,
\be
\label{eq:trasf2m}
A^{''}_{-} = |\mu| A'_{-0} - |\nu|  A^{'*}_{-0}\ ,
\ee
where we have defined
\ba
\label{eq:bmdem2m}
A^{'}_{-0} &\equiv& e^{i \lt \fr{\theta_\mu - \theta_\nu}{2} \rt}  A_{-0}\ , \\
A^{''}_{-}  &\equiv& e^{-i \lt \fr{\theta_\mu+ \theta_\nu }{2}\rt} A_{-}\ .
\ea
This will cause the amplification to occur along the $Y$ quadrature for mode $-$, and renders in this sense Eq.(\ref{eq:trasf2m}) different from the canonical form of Ref. \cite{Caves_82}.
With a definition of the quadratures for the fields $A^{'}_\pm$ and $A^{''}_\pm$ analogous to the one in Eq. (\ref{eq:def_quadr}) we obtain from Eqs. (\ref{eq:trasf2p}) and (\ref{eq:trasf2m})
\ba
\label{eq:amp_diago3}
X^{''}_{\pm} &=&  (|\mu| \pm  |\nu|) X'_{\pm0}\ , \\
Y^{''}_{\pm} &=&  (|\mu| \mp |\nu|) Y'_{\pm0}\ .
\ea
We thus obtain the decomposition
\ba
\label{eq:b-mvrb}
M_{\pm} &=& C_\pm U  \Sigma_\pm  W^T \nn \\
&=& C_\pm \left( 
\begin{array}{cccccccc}
\cos \theta & - \sin \theta \\
\sin \theta & \cos \theta   \\
\end{array} 
\right)  \hspace{-0.1cm} \left( 
\begin{array}{cccccccc}
|\mu| \pm  |\nu|& 0 \\
0 & |\mu| \mp  |\nu|  \\
\end{array} 
\right)  \nn\\
&&\times
 \hspace{-0.1cm} \left( 
\begin{array}{cccccccc}
\cos \phi &  \sin \phi \\
-\sin \phi & \cos \phi   \\
\end{array} 
\right),  
\ea
where
$C_\pm = M_\pm W \Sigma_\pm^{-1} U^T $, and where from Eqs. (\ref{eq:bmdem2p}) and (\ref{eq:bmdem2m}) the angles $\theta$ and $\phi$ result in the same definitions as given in Eqs.(\ref{eq:sol_angles3}), (\ref{eq:sol_angles4}). 


\begin{figure} []
\includegraphics[width=0.9\columnwidth]{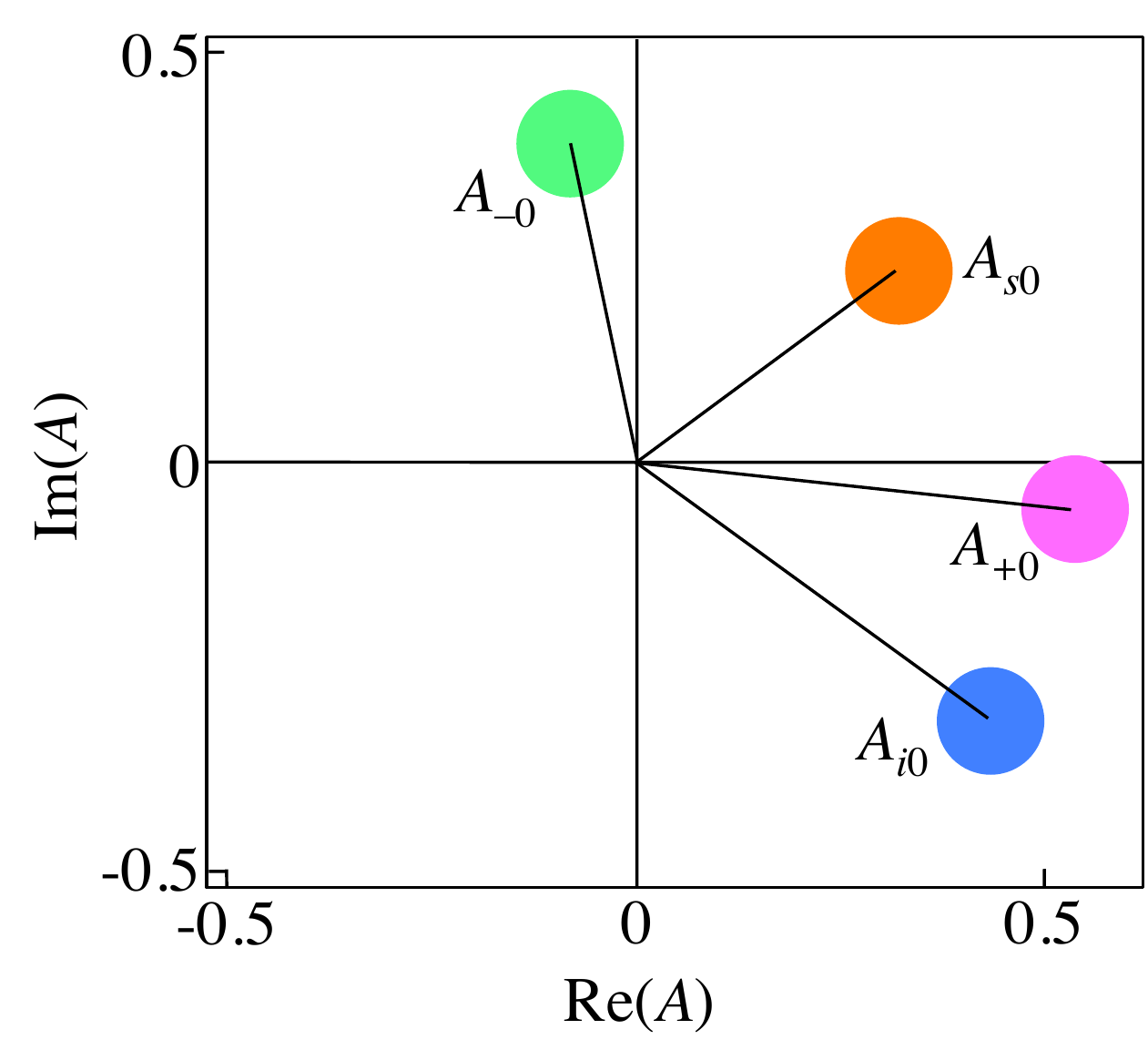}
\caption{Phasor representation for the input fields in the case of configuration ``B''. The units for the quadratures are arbitrary. The incident signal and idler fields $A_{s0}$ and $A_{i0}$, which are in coherent states  $|\alpha_{s0}\rangle$ with $\alpha_{s0}=0.4 \exp{ (i \pi /5)}$ and $|\alpha_{i0}\rangle$ with $\alpha_{i0}=0.54 \exp{ (-i \pi /5)}$, respectively, are transformed into the fields $A_{+0}$ and $A_{-0}$ according to Eqs. (\ref{eq:changing1}) and (\ref{eq:changing2}).}\label{Fig03} 
\end{figure}
\begin{figure} [h!]
\includegraphics[width=1.0\columnwidth]{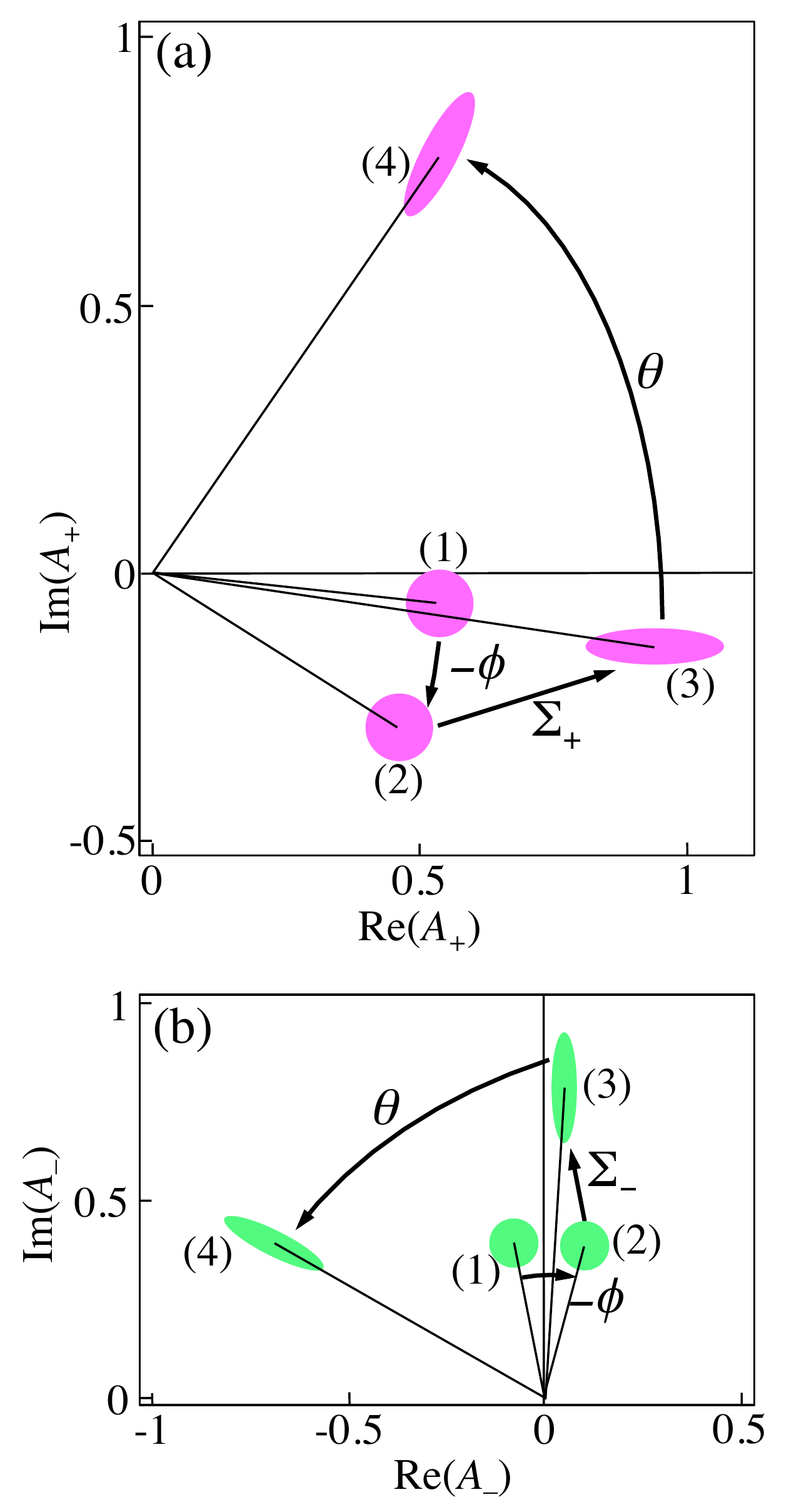}
\caption{Evolution of the `+' (a) and `-' (b) fields in the phasor representation along the three steps of the Bloch-Messiah decomposition of the parametric amplification in configuration ``B''. The units for the quadratures are arbitrary. The incident coherent fields $A_{\pm0}$ (1), which were already plotted in Fig. \ref{Fig03}, undergo a first rotation by an angle $-\phi$, resulting in field (2). These fields are transformed into field (3) by the squeezing operators $\Sigma_{\pm}$, before being rotated by an angle $\theta$ to give the output fields $A_{\pm}$ (4). The calculations were performed for a fiber length $z=300\ \mathrm{m}$, $\gamma=11.3\times 10^{-3}\,\mathrm{W}^{-1} \mathrm{m}^{-1}$, $\Delta\beta= -4.54\times10^{-11}\,\mathrm{m}^{-1}$, $P_2= 230\,\mathrm{mW}$.}\label{Fig04}
\end{figure}

\begin{figure} [h]
\includegraphics[width=0.9\columnwidth]{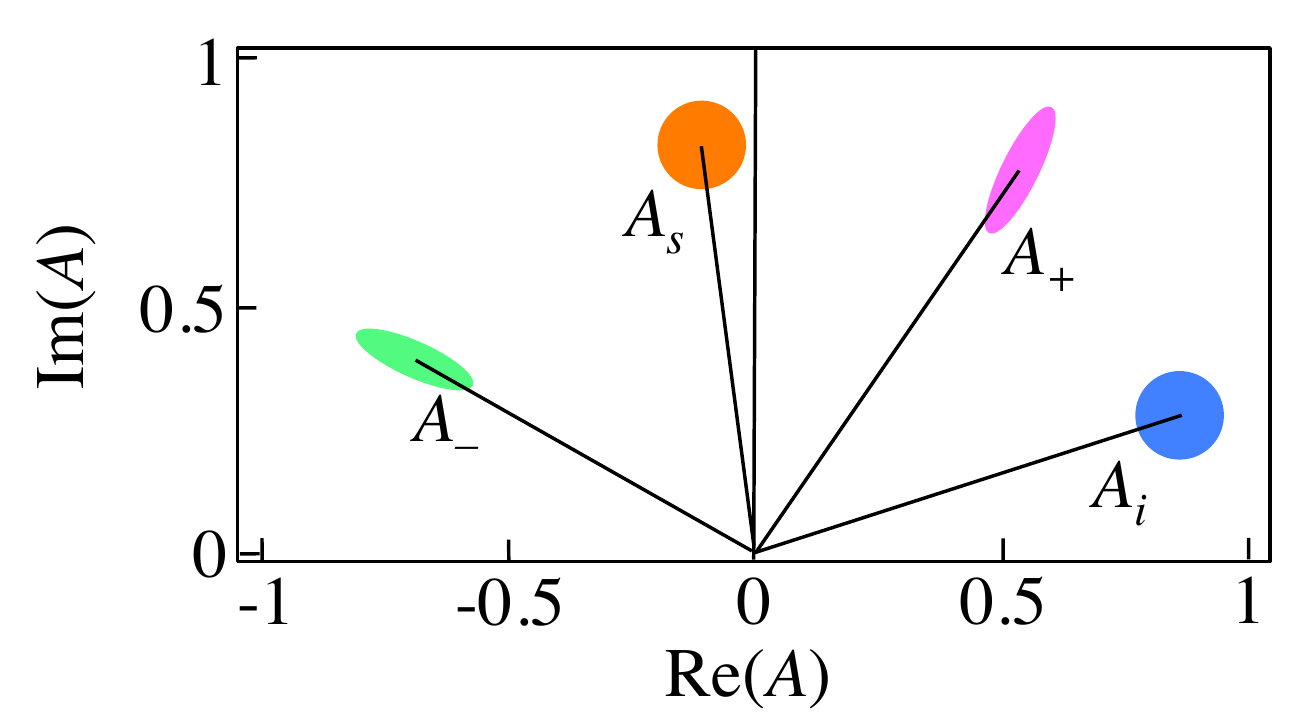}
\caption{Phasor representation for the output fields in the case of configuration ``B''. The units for the quadratures are arbitrary. The output signal and idler fields $A_{s}$ and $A_{i}$ are obtained from the output fields $A_{+}$ and $A_{-}$, which were obtained in Fig.\ \ref{Fig04}, using Eqs.\;(\ref{eq:changing1}) and (\ref{eq:changing2}).}\label{Fig05}
\end{figure}

An example of application of this formalism is shown graphically in Figs.\ \ref{Fig03}, \ref{Fig04}, and \ref{Fig05}. Fig.\ \ref{Fig03} illustrates the application of Eqs.\ (\ref{eq:changing1}) and (\ref{eq:changing2}): the fields $A_{+0}$ and $A_{-0}$ are obtained from the input signal and idler fields $A_{s0}$ and $A_{i0}$, which are supposed here to be in a coherent state. 
Analogously as it was for the signal field in configuration ``A" in Sec.\ref{se:bloch-messiah-configA}, each of these input fields are independently transformed through the three steps of the Bloch-Messiah decomposition of their transfer matrix, as given in Eq.\ (\ref{eq:b-mvrb}). We took the same  values for the fiber parameters as in the case ``A'' (see Fig. \ref{Fig02} and accompanying text). In this calculation we suppose that the pump wavelength is located at the zero dispersion wavelength of the fiber (1547 nm in this case) and that the signal and idler are located at $\pm0.32\,\mathrm{nm}$ from the pump. Eqs.\ (\ref{eq:sol_angles3}) and (\ref{eq:sol_angles4}) lead to angles $\phi = -26.0\,^{\circ}$ and $\theta = 64.0\,^{\circ}$ for the transformation of the `+' and `-' fields. The amplitude gain coefficient is equal to $|\mu|+|\nu|=2.05$. The way these parameters can be directly derived from the experimental parameters is developed in Appendix \ref{AppendixD}.

Finally, Fig.\ \ref{Fig05} shows the application of Eqs.\ (\ref{eq:changing1}) and (\ref{eq:changing2}) permits to retrieve the output signal and idler fields $A_{s}$ and $A_{i}$ from the fields $A_{+}$ and $A_{-}$ that were determined in Fig.\ \ref{Fig04}.

One can notice from Fig.\ \ref{Fig05} that in this configuration, the noise ellipses for the `+' and `-' modes are orthogonal. The fluctuations of signal and idler, though, are isotropic, as we will quantitatively show in Sec.\ref{se:snrB}.
Indeed, one should not conclude from Fig.\ \ref{Fig05} and from Eqs.\ (\ref{eq:changing1}) and (\ref{eq:changing2}) that the modes `+' and `-' on the one hand, and the idler and signal modes on the other hand, play symmetric roles. The orthogonal eigenmodes of our symplectic transformation are the `+' and `-' modes (and not the signal and idler modes). This leads to the fact that, in general, the signal and idler modes are entangled at the output of the amplifier. This point as well will appear clearly when we write the covariance matrix in the two basis in Sec.\ref{se:snrB}.

\section{Optimum input condition} 

Based on the Bloch-Messiah decomposition given and illustrated in the preceding section, Figs.\ \ref{Fig02} to \ref{Fig05} show that the gain experienced by the signal and/or the idler depends on the quadrature along which the fields are injected. The aim of this Section is thus to derive and illustrate the conditions for which the gain is optimized.

\subsection{Configuration ``A''}
\label{se:config_A_opt_cond}

In the case where one is interested in amplifying an input coherent state with a maximum gain without any added noise, one must have the field in the amplification step aligned with the  eigen-quadrature corresponding to $|\mu|+|\nu|$, i. e. the largest of the two eigenvalues of the matrix $\Sigma$ [see Eq.\;(\ref{eq:b-mvra})]. By looking at Fig.\;\ref{Fig02}, this means that the first rotation of angle $-\phi$ of the Bloch-Messiah decomposition must bring the input coherent field along the horizontal axis. Hence, if the input field is $A_{s0} = |A_{s0}| \mathrm{e}^{\mathrm{i} \theta_{s0}}$
(or equivalently, in a quantum formalisms, if it is in a coherent state $|\alpha_{s0}\rangle$ with $\alpha_{s0}=A_{s0}$), one has to choose $\theta_{s0}$ such that
\ba
 \theta_{s0} = \frac{\theta_{\nu} - \theta_{\mu} }{2} = - \phi\ , \label{eq53}
\ea
with $\phi$ as in Eq.(\ref{eq:sol_angles4}) \cite{Footnote}. In mathematical terms, substituting condition (\ref{eq53}) into Eq. (\ref{stdeqtris}) yields
\ba
\label{stdeqtris22}
A_s \mathrm{e}^{\mathrm{i} \lt \fr{\theta_\mu - \theta_\nu}{2} \rt} = |A_{s0} | ( |\mu |+ |\nu| )\ ,
\ea
showing that the full mean field of the signal is amplified and not only a projection along one axis. 

\begin{figure} [h]
\includegraphics[width=0.9\columnwidth]{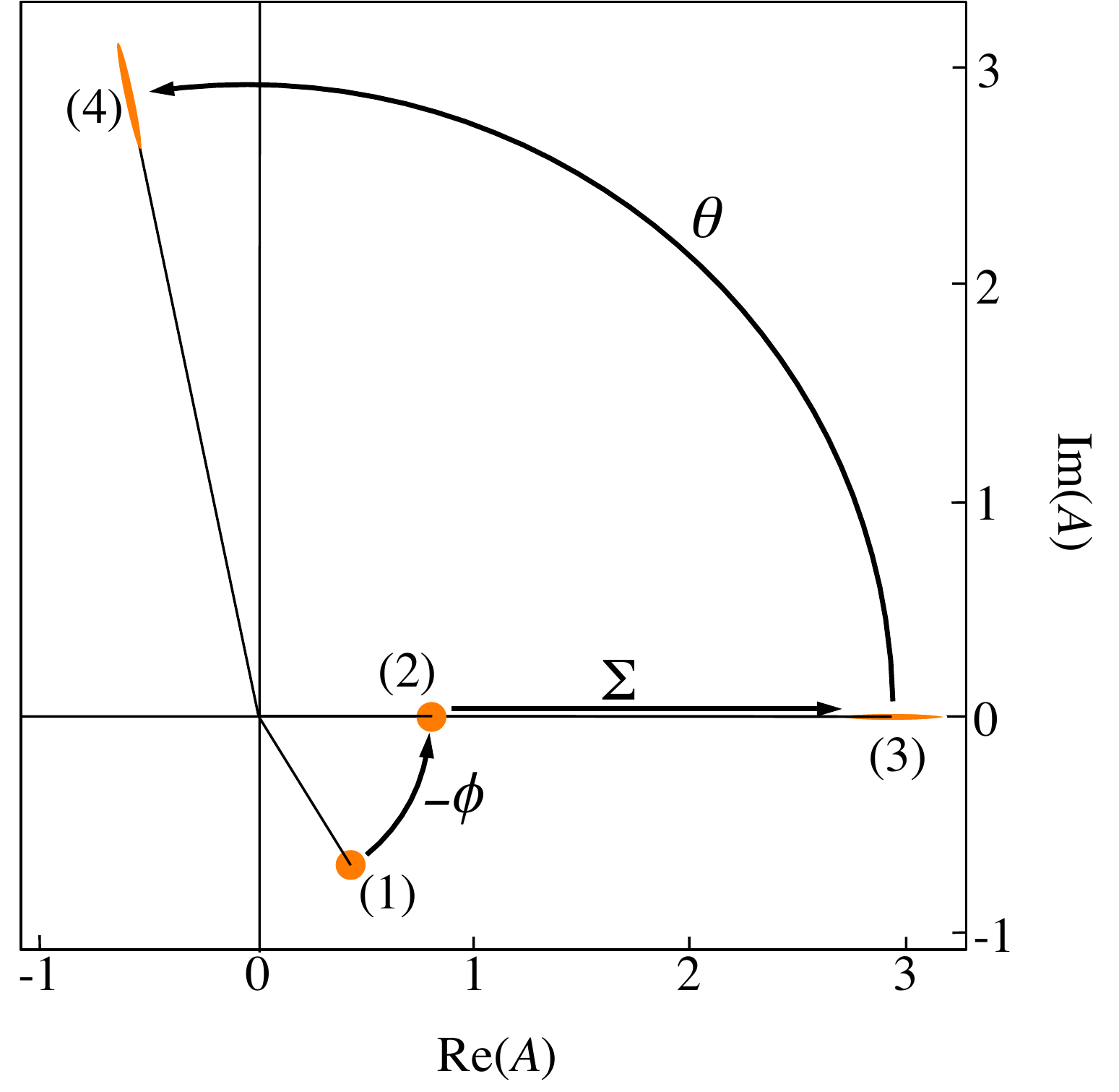}
\caption{Evolution of the field in the phasor representation along the three steps of the Bloch-Messiah decomposition of the parametric amplification in configuration "A" in the case of optimal coupling. All parameters are the same as in Fig.\ \ref{Fig02}, except for the phase of the input field which is chosen according to Eq.\ (\ref{eq53}).}\label{Fig06}
\end{figure}

Fig.\ \ref{Fig06} is an illustration of such an optimal input coupling. Compared with the situation of Fig.\ \ref{Fig02}, the only change that has been made is in the choice of the angle of the input quadrature. The mean value of the field and the fiber and pump parameters are identical to those of Fig.\ \ref{Fig02}. The result of the choice of the optimum quadrature is that the output field undegoes the maximum amplification. As it can be seen from Eq.(\ref{eq:clax_gain}), this corresponds to the largest classical gain, equal in this case to $G_{\text{max}}=(|\mu|+|\nu|)^2=13.3$. In this case, the fluctuation ellipse is aligned along the orientation of the mean value of the field, contrary to the case of Fig.\ \ref{Fig02}.

\subsection{Configuration ``B''}
\label{se:config_B_opt_cond}

In the case where signal and idler are no longer degenerate, a similar optimum input condition can be considered, except now, as can be seen from Fig.\ \ref{Fig04}, it has to be considered for the modes `+' and `-', which are combinations of the signal and idler modes. In this case, the information to be carried has to be encoded in both signal and idler modes. To be fully convinced of this point, consider first the case in which only one mode, say the signal, contains the information in its input state, i.e. $A_{i0} = 0$. The same discussion can be carried out for quantum fields choosing an initial vacuum state in the idler mode, as will be done in Sec.\ref{se:snrB}.

One then obtains from Eq.(\ref{solA1}):
\ba
A_s &=& \mu A_{s0}\ ,\label{eq:PIA1}\\
  A_i &= &\nu A_{s0}^*\ .\label{eq:PIA2}
\ea
Eqs.\ (\ref{eq:PIA1}) and (\ref{eq:PIA2}) correspond to a phase insensitive amplifier, with gain
\be
G_{PIA} = \fr{P_s}{P_{s0}} = |\mu |^2,
\ee
and hence effectively not useful for noiseless amplification with a gain larger that $1$~\cite{Caves_82}.

Let us now consider instead the case in which both signal and idler are present at the input, with the same power and a specific phase relation.
Following the same graphical argument as the one presented in Sec. \ref{se:config_A_opt_cond}, we might guess that the optimal condition must be the one for which the fields $A_+$ and $A_-$ are brought onto the $x$ and $y$ axis respectively after the first rotation of the Bloch-Messiah decomposition. 
Rewriting Eq. (\ref{solA1})
\ba
A_s = |\mu | \mathrm{e}^{\mathrm{i} \theta_\mu} A_{s0} + |\nu| \mathrm{e}^{\mathrm{i} \theta_\nu} A^*_{i0}\ , \label{new_eq1}\\
A_i = |\nu| \mathrm{e}^{\mathrm{i} \theta_\nu} A_{s0}^* + |\mu| \mathrm{e}^{\mathrm{i} \theta_\mu} A_{i0}\ ,\label{new_eq2}
\ea
we see that in order to reach maximal gain we have to set $\mathrm{e}^{\mathrm{i} \theta_\mu} A_{s0} = \mathrm{e}^{\mathrm{i} \theta_\nu} A^*_{i0}$, i.e.
\be
\label{magin_cond}
A_{i0} = A_{s0}^*  \mathrm{e}^{- \mathrm{i} (\theta_\mu - \theta_\nu)}.
\ee
Under the condition Eq.(\ref{magin_cond}) we obtain indeed from Eqs. (\ref{new_eq1}) and (\ref{new_eq2})
\ba
A_s \mathrm{e}^{-\mathrm{i} \theta_\mu}& = &( |\mu | + |\nu|) A_{s0}\ , \label{stdeqtris2-1}\\
A_i \mathrm{e}^{-\mathrm{i} \theta_\mu} &=& ( |\mu | + |\nu|)  A_{i0}\ ,\label{stdeqtris2-2}
\ea
i.e., the amplitude of both  signal and idler is  with gain $|\mu | + |\nu|$.
Note that condition (\ref{magin_cond}) implies the phase relation
\be
\label{eq:tong}
\theta_\mu - \theta_\nu + \theta_{s0} + \theta_{i0} = 0  \mbox{   mod  } 2 \pi,
\ee
which is precisely the condition maximizing the power gain in Eq. (\ref{eq:gps2_maxmin}) (the parameter $\eta$ being obviously equal to one since signal and idler have equal powers). Graphically, one can see from Fig.\ \ref{Fig04} that this condition corresponds to the situation in which the input fields in modes `+' and `-' are brought onto the horizontal and vertical quadratures, respectively, after the first rotation of the Bloch-Messiah decomposition, thus maximizing their gains, as expected. The condition (\ref{eq:tong}) for optimal amplification has been provided in Ref.~\cite{Tong_11}. There, Eqs.(\ref{magin_cond}) and (\ref{eq:tong}) are experimentally implemented by feeding the amplifier with the fields coming out of a copier stage.

\begin{figure} []
\includegraphics[width=0.9\columnwidth]{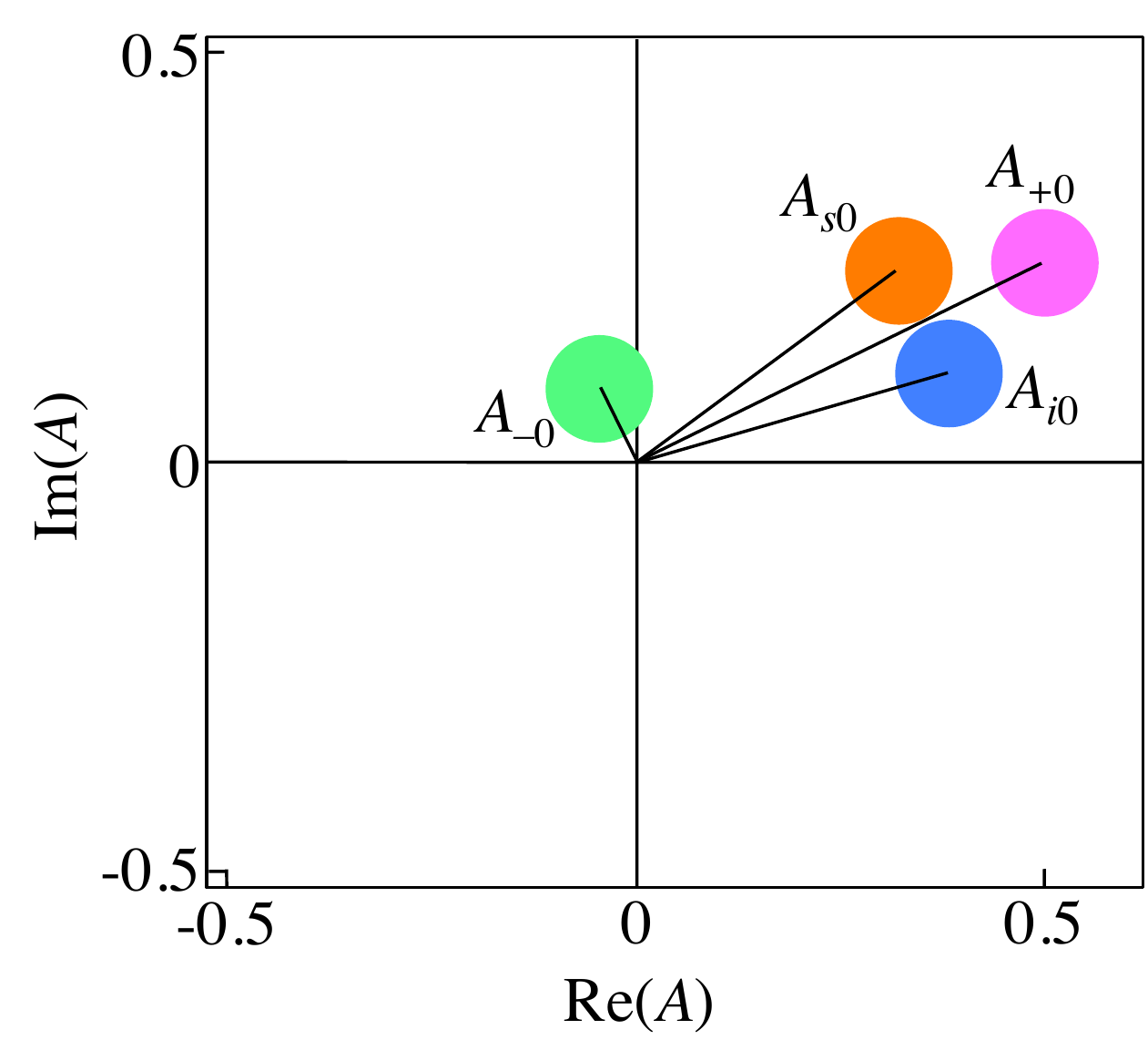}
\caption{Phasor representation for the input fields in the case of configuration ``B'' for optimal coupling. The units for the quadratures are arbitrary. All parameters are the same as in Fig.\ \ref{Fig03}. $\alpha_{s0}=0.4 \exp{ (i \pi /5)}$ like in Fig.\ \ref{Fig03}.  Only the idler input field is different from Fig.\ \ref{Fig03}: it is chosen according to Eq.\ (\ref{magin_cond}). } \label{Fig07}
\end{figure}

\begin{figure} [h]
\includegraphics[width=1.0\columnwidth]{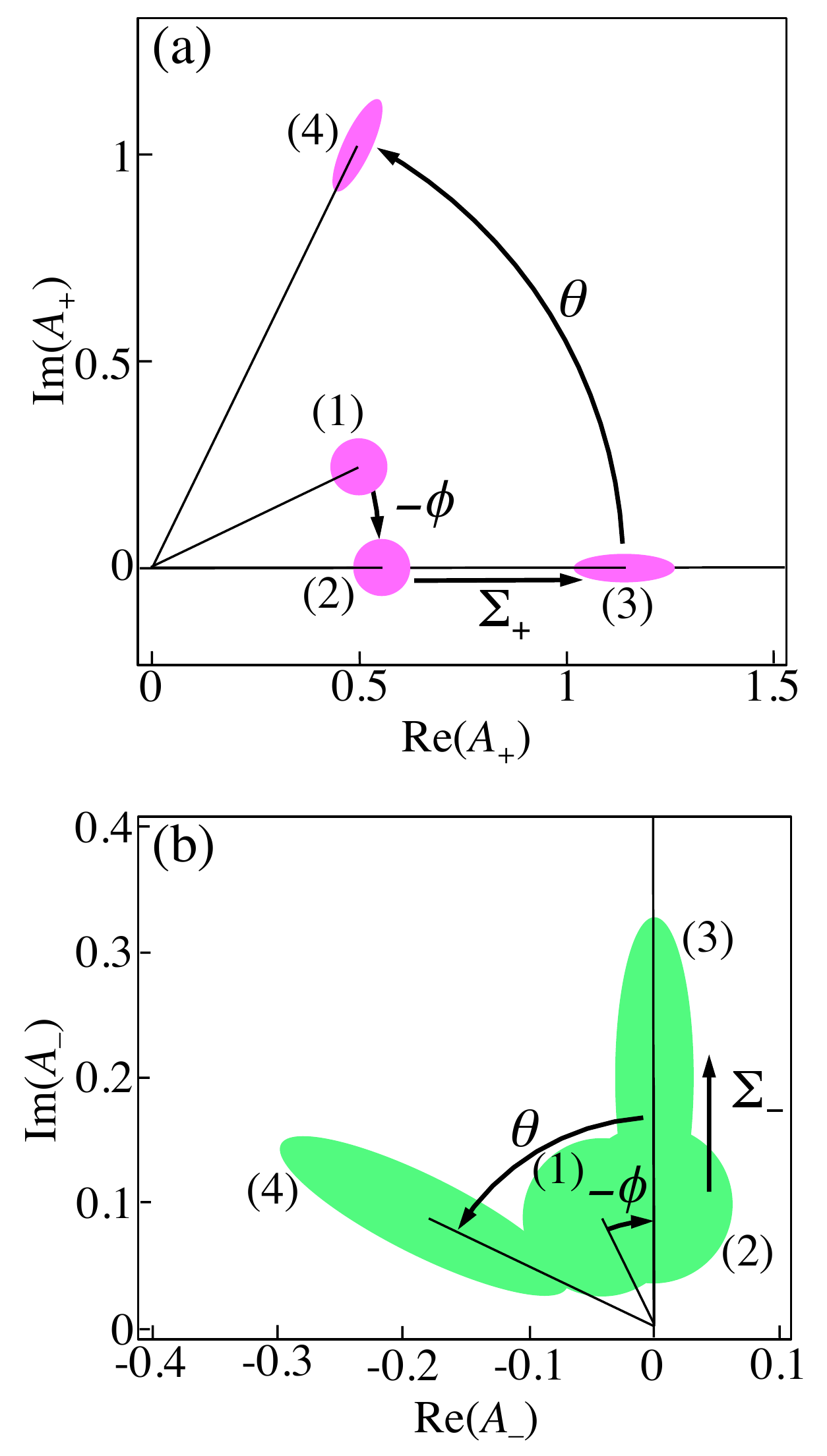}
\caption{Same as Fig.\ \ref{Fig04} for the optimum input conditions of Fig.\ \ref{Fig07}.}\label{Fig08}
\end{figure}

\begin{figure} [h]
\includegraphics[width=0.9\columnwidth]{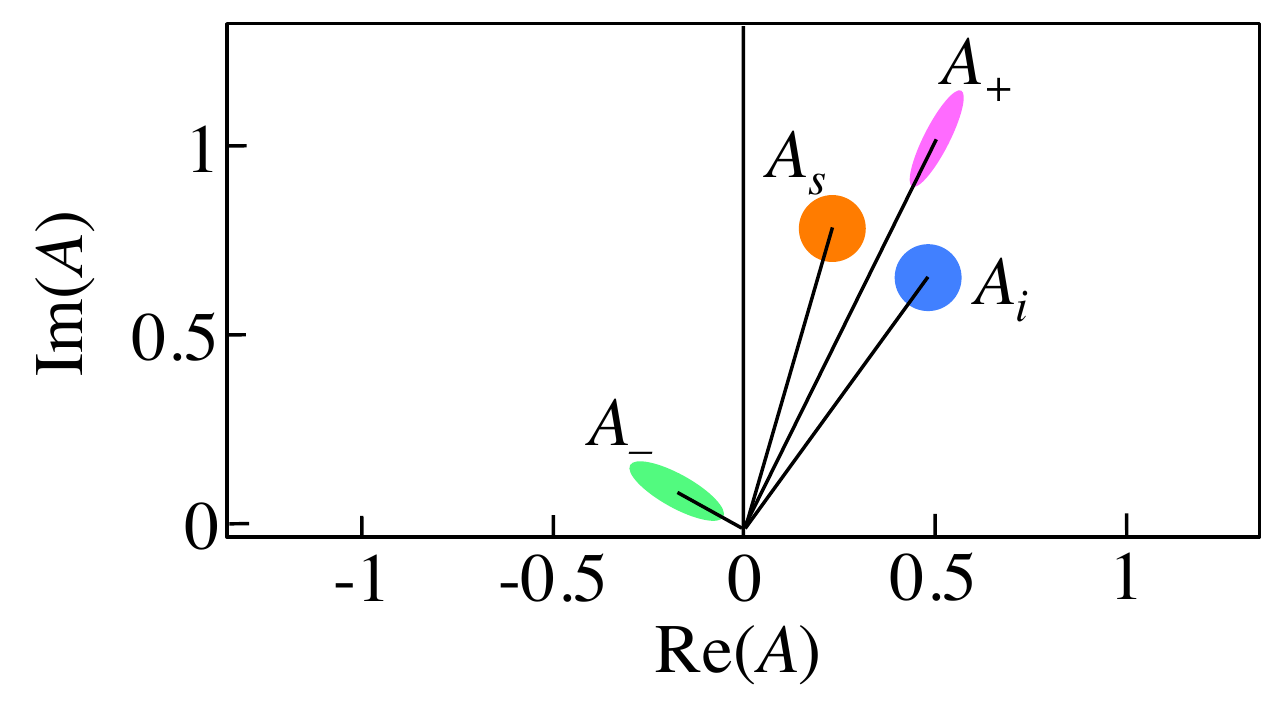}
\caption{Same as Fig.\ \ref{Fig05} for the optimum input conditions of Fig.\ \ref{Fig07}. }\label{Fig09}
\end{figure}

An illustration of this optimum injection condition is given in Figs.\ \ref{Fig07}-\ref{Fig09}, which have been obtained for the same parameters as Figs.\ \ref{Fig03}-\ref{Fig05}, with the same input signal field, except now the idler field is chosen according to Eq.\ (\ref{magin_cond}). 

From Eqs. (\ref{stdeqtris2-1}) and (\ref{stdeqtris2-2}), we see that for both signal and idler, with this special input condition, the effective amplification takes place along the direction of the respective mean fields, with maximum gain  $|\mu|+|\nu|$ here equal to 2.05 with our parameters (see Fig.\ \ref{Fig09}).
Note that the signal and idler fields, contrarily to the modes $+$ and $-$, are correlated at the output. The correlations between signal and idler at the output of the amplifier are addressed in Section \ref{se:snrB}.

\se{Noise Figure and Correlation Calculations}

In this section we present the calculation of the amplifier noise-figure (or figure of merit) in the case of a homodyne detection, for both configurations ``A'' and ``B''. We use a quantum formalism, replacing all capital c-numbers such as $A_s$ and $A_i$ for example, by their quantum operator counterparts $\hat a_s$ and $\hat a_i$. In the case of configuration ``B'', the fact that we deal with two quantum fields leads to interesting predictions concerning the correlations between these fields \cite{Sharping2001} at the output of the amplifier which are detailed in the second subsection below.

\subsection{Configuration ``A'': Noise Figure}
\label{se:snrA}

 In order to properly study the amplifier noise figure, we chose to compare the signal in input and output as they could be measured by a homodyne detection.
This measurement has been shown to be able to reach the fundamental limits in terms of extractable information~\cite{Pinel_12}.

We start by computing the signal-to-noise ratio (SNR) in input and in output. Consider an input coherent state $|\alpha_{s0} \rangle$.
We write the generic quadrature as $\hat x_\varphi =( \hat a e^{-i \varphi} + \hat a^{\dagger} e^{i \varphi})/2$, consistently with the definition in Eq.\,(\ref{def_quadratures1}), which also fixes the convention $\Delta^2 {x_\varphi}_{vac} = 1/4$ for the vacuum fluctuations. In order to properly quantify the input signal power the only meaningful choice is $\varphi_0 = \theta_{s0}$, and we trivially obtain the signal
\ba
\langle \hat x_{s \theta_{s0}} \rangle & =& ( \alpha_{s0} e^{-i \theta_{s0}} +  \alpha_{s0}^* e^{i \theta_{s0} })/2 =  |\alpha_{s0}|  ,\label{Eq74}
\ea
where we have that $\langle  \alpha |\hat a | \alpha \rangle = \alpha$~\cite{Haroche}, 
which corresponds to the standard vacuum fluctuations
\be
\Delta^2 {x_{s \varphi_0}} = \fr{1}{4}\ , \label{Eq75}
\ee
and where we have simplified the notations by identifying $\hat x_{s \theta_{s0}} \equiv \hat x_{s  \theta_{s0},0}$.
We obtain hence for the input signal-to-noise ratio
\be  
\label{eq:input_snr}
\mathrm{SNR}_{\text{in}} = \fr{\langle \hat x_{s  \theta_{s0}} \rangle^2}{\Delta^2 {x_{s  \theta_{s0}}}} = 4  |\alpha_{s0}|^2. 
\ee
The corresponding output quantities are easily obtained using Eq.(\ref{eq:matrici}), leading to the following value for the expected outcome of any output field quadrature:
\be
\label{eq:823}
\langle \hat x_{s, \varphi} \rangle = |\alpha_{s0}| [ |\mu| \cos(\theta_\mu+ \theta_{s0} - \varphi)+ |\nu| \cos(\theta_\nu- \theta_{s0} - \varphi)],
\ee
with noise (see Appendix \ref{app_NF})
\be
\Delta^2 {x_{s \varphi}}  =  \fr{ 2 |\mu | |\nu | \cos (\theta_\mu + \theta_\nu - 2 \varphi) + |\mu |^2 + |\nu |^2 }{4}.\label{Eq78}
\ee
Hence we obtain the output signal-to-noise ratio:
\ba
 & & \mathrm{SNR}_{\text{out}} = \nn \\
 & & \fr{4 |\alpha_{s0}|^2 [ |\mu| \cos(\theta_\mu+ \theta_{s0} - \varphi)+ |\nu| \cos(\theta_\nu- \theta_{s0} - \varphi)]^2}{ 2 |\mu | |\nu | \cos (\theta_\mu + \theta_\nu - 2 \varphi) + |\mu |^2 + |\nu |^2 }.\nn \\
 & & \label{Eq79}
\ea
The corresponding noise figure is thus obtained by combining Eqs.\ (\ref{eq:input_snr}) and (\ref{Eq79}):
\ba
\label{eq:NFnolossA}
\mathrm{NF} &=& \fr{\mathrm{SNR}_{\text{in}}}{\mathrm{SNR}_{\text{out}}} \nn \\
&=& \fr{\lqu 2 |\mu | |\nu | \cos (\theta_\mu + \theta_\nu - 2 \varphi) + (|\mu |^2 + |\nu |^2 )\rqu}{[ |\mu| \cos(\theta_\mu+ \theta_{s0} - \varphi)+ |\nu| \cos(\theta_\nu- \theta_{s0} - \varphi)]^2}. \nn \\
& &
\ea
It is easily seen that under optimum input condition given in Eq. (\ref{eq53}) one obtains $\mathrm{NF} = 1$ if the detection is performed along the quadrature phase $\varphi = \theta = (\theta_\mu + \theta_\nu)/2$.  This appears natural since this is the phase of the output field as can be visualized Fig. \ref{Fig06}. Hence, there is a detection phase for which the signal to noise ratio is preserved by the amplifier, consistently with the fact that in ideal conditions, i. e., in the absence of losses, the phase sensitive amplifier can amplify without adding any excess noise.
Eq. (\ref{eq:NFnolossA}) coincides with the result of Eq. (86) in Ref.~\cite{McKinstrie_05} for the case of optimal local oscillator phase. Note that in the special case of the optimum input condition an intensity detection also allows to recover a unit noise figure due to the fact that the output state is phase squeezed, i.e. the fluctuation ellipse is aligned with the mean field. (see Fig. \ref{Fig06})

\subsection{Configuration ``B'': Noise Figure and Generation of Correlated Photons}
\label{se:snrB}

\subsubsection{Noise Figure}

  We derive now the figure of merit of the amplifier in the second configuration, assuming that in input there is a coherent state (arbitrary for the moment) in both the signal and the idler modes, i.\,e.  $|\psi_{0} \rangle = |\alpha_{s0} \rangle |\alpha_{i0} \rangle$. 
In this case, in order to properly quantify the input information we have to consider the mode quadrature which resumes the full signal and idler amplitudes, namely 
\be
\hat x_{0} = \fr{\hat x_{s \theta_{s0}} + \hat x_{s \theta_{i0}}}{\sqrt{2}}.
\ee
For simplicity consider the specialization to the case of the optimum input condition expressed by Eq.\,(\ref{eq:tong}), jointly with $P_s = P_i$.
Then 
\be
\langle \hat x_{0} \rangle = \fr{|\alpha_{s0}| +  |\alpha_{i0}| }{\sqrt{2}} = \sqrt{2} |\alpha_{s0}|.
\ee
The corresponding fluctuations are
\be
\Delta^2  x_{0} = \fr{\Delta^2  x_{s \theta_{s0}} + \Delta^2  x_{i \theta_{i0}} }{2} = \fr{1}{4}
\ee
yielding the input signal-to-noise ratio 
\be
\label{eq:input-snrB}
\mathrm{SNR}_{s,\text{in}} = \fr{\langle \hat x_{0} \rangle^2}{\Delta^2  x_{0}} = 8  |\alpha_{s0}|^2 = 8 P_s.
\ee
Analogously, we consider in output 
\be
\label{eq:output_amp9}
\langle \hat x \rangle =  \fr{\langle \hat x_{s \theta_{s}}  \rangle +  \langle \hat x_{s \theta_{i}}  \rangle}{\sqrt{2}} = \sqrt{2} ( |\mu | + |\nu|) |\alpha_{s0}| ,
\ee
where $\langle \hat x_{s \theta_{s}} \rangle = ( |\mu | + |\nu|)  |\alpha_{s0}|$ and $\langle \hat x_{i \theta_{i}} \rangle = ( |\mu | + |\nu|)  |\alpha_{i0}|$ can be obtained with the help of Eqs.\ (\ref{stdeqtris2-1},\ref{stdeqtris2-2}). From the same equation set one can conclude that the relevant output phases in which to perform the homodyne  detection are $\theta_{s} = \theta_\mu + \theta_{s0}$ and $\theta_{i} = \theta_\mu + \theta_{i0} = \theta_\nu + \theta_{s0}$ (see also Fig.~\ref{Fig08}).
The corresponding output fluctuations are derived in Appendix \ref{app:fluctuations_mode_output} and result in 
\be
\label{eq:flutt_mode_interest}
\Delta^2  x = \fr{( |\mu | + |\nu|)^2}{4},
\ee
yielding an output signal-to-noise ratio
\be
\mathrm{SNR}_{s,\text{out}} = \fr{\langle \hat x \rangle^2}{\Delta^2  x} = 8 |\alpha_{s0}|^2 = 8 P_s.
\ee
This obviously leads by comparison with Eq.\,(\ref{eq:input-snrB}) to the unit noise figure 
\be
\mathrm{NF} =  \fr{\mathrm{SNR}_{\text{in}}}{\mathrm{SNR}_{\text{out}}} = 1,
\ee
which shows that with the good choice of the detection mode and phase it is possible to recover in output all the input information with no degradation of the signal-to-noise ratio.

It is interesting to consider the noise figure obtained by measuring in output the signal only.
The corresponding output signal reads
$\langle \hat x_{s\varphi} \rangle =  ( |\mu|  + |\nu| )|\alpha_{s0}|$ with noise $\Delta^2 {x_{s \varphi}} =\fr{ |\mu|^2  + |\nu|^2}{4}$, as will be obtained later in Eq.\,(\ref{eq:fluttuazioni}). This leads to the output signal-to-noise ratio
\ba
\mathrm{SNR}_{s,\text{out}} =  \fr{4( |\mu| + |\nu| )^2 |\alpha_{s0}|^2}{( |\mu|^2  + |\nu|^2)}  \label{eq:hdkkjehd}\ , 
\ea 
where we have chosen the optimal detection phase $\varphi = \theta_\mu + \theta_{s0}$.
Combining Eq.\,(\ref{eq:hdkkjehd}) with the input signal-to-noise ratio of Eq.\,(\ref{eq:input-snrB}) we obtain the noise figure 
\ba
\label{eq7}
\mathrm{NF}_s  = \fr{\mathrm{SNR}_{s, \text{in}}}{\mathrm{SNR}_{s, \text{out}}} = \fr{2 ( |\mu|^2  + |\nu|^2)}{( |\mu| + |\nu| )^2 } \hspace{0.1cm} \stackrel{\footnotesize{\mbox{large  gain}}}{\rightarrow}   \hspace{0.1cm}   = 1,
\ea
where the arrow indicates the limit $ |\mu|^2, |\nu|^2 \gg 1$ where $|\mu|^2 \simeq |\nu|^2$ due to $|\mu|^2 - |\nu|^2 = 1$~\cite{Vasiljiev, radic2005, Tong_10}.
Eq.\,(\ref{eq7}) shows that in the large gain limit one can recover all the information in output by measuring the signal only, despite the fact that the output idler power is non-zero. This apparent paradox is explained by the fact that signal and idler become in this limit highly correlated, as will appear clear in the forthcoming section.  
This also explains the negative noise figure obtained in some papers considering signal-to-noise ratios with respect to the signal only, both in input and in output. This obviously does not take into account the input idler power, and leads for general input fields and output detection phase to~\cite{radic2005, Vasiljiev}
\ba
\label{eq:NFnolossB}
& & \mathrm{NF}_s =  \\ 
& & \fr{ |\alpha_{s0} |^2  ( |\mu |^2 + |\nu |^2)}{ \lqu | \mu |  |\alpha_{s0} | \cos (\theta_\mu + \theta_{s0} - \varphi) + | \nu |  |\alpha_{i0} | \cos (\theta_\nu - \theta_{i0} - \varphi) \rqu^2}, \nn 
\ea
yielding $\simeq 1/2 \simeq - 3$dB for the optimal input condition.

Also note that using different conditions at the input of the amplifier, e.g. taking the vacuum as input in the idler mode, yields less favorable results. Considering for instance the vacuum in input of the idler mode $ |0_{i0} \rangle $ one has for $\varphi = \theta_\mu + \theta_{s0}$
\be
\langle \hat x_{s\varphi} \rangle =  |\mu| \langle \hat x_{s0} \rangle =  |\mu|   |\alpha_{s0} | \cos(\theta_{s0} - \varphi),
\ee
yielding the PIA figure of merit
\be
\label{eq12}
\mathrm{NF} = \fr{\mathrm{SNR}_{\text{in}}}{\mathrm{SNR}_{\text{out}}} =   \fr{ |\mu|^2  + |\nu|^2}{|\mu|^2}  \hspace{0.1cm} \stackrel{\footnotesize{\mbox{large  gain}}}{\rightarrow}     \hspace{0.1cm}  \fr{2 |\mu|^2}{ |\mu|^2} = 2 \simeq 3 \text{dB},
\ee
where extra noise has been introduced from the vacuum fluctuations of the idler mode.
This is consistent with what found in Sec.\ref{se:config_B_opt_cond}. The noise figure for direct detection (i.e., related to the current) yields in these special cases the same results (see e.g. Ref.~\cite{McKinstrie_05}).

\subsubsection{Generation of quantum correlated photons}

Beyond the characterization of the noise properties of the amplifier output presented above, we characterize here the correlations between the signal and idler output fields.
A compact way of evaluating the fluctuations and correlations of the signal and idler modes is to evaluate their full covariance matrix.
As a starting point, we assume that we have a coherent state in input (possibly the vacuum) for the signal mode, and one for the idler. Then the input covariance matrix expressed in the basis $\{ \hat  x_s,\hat x_i,\hat y_s,\hat y_i \}$ reads $\Sigma_{si,0} = \mathcal{I}/4$, where we use the convention $\Delta^2_{\text{vac}} = \langle \hat x^2 \rangle_{\text{vac}} = \langle \hat y^2 \rangle_{\text{vac}} = 1/4$ and where $\mathcal{I}$ is the $4\times 4$ identity matrix. $\hat  x_s,\hat x_i,\hat y_s,\hat y_i$ are the operators corresponding to the c-numbers $X_s,X_i,Y_s,Y_i$ defined in the same manner as in Eq. (\ref{def_quadratures1}) .

Then, we apply the two-mode symplectic transformation which describes the modes evolution on the fields quadratures. This can be directly derived from Eqs. (\ref {solA1}) and (\ref {solAbis}), yielding
\ba
\label{eq:trasf_modes}
\left( 
\begin{array}{cccccccc}
\hat{x}_{s} \\  
\hat{x}_{i}   \\
\hat{y}_{s} \\ 
\hat{y}_{i}  \\
\end{array} 
\right) \hspace{-0.13cm} &=& \hspace{-0.13cm}
\left( 
\begin{array}{cccccccc}
\mathcal{R}e[\mu] & \mathcal{R}e[\nu] & -\mathcal{I}m[\mu]  & \mathcal{I}m[\nu] \\  
\mathcal{R}e[\nu] & \mathcal{R}e[\mu]  &  \mathcal{I}m[\nu]  & -\mathcal{I}m[\mu]  \\ 
\mathcal{I}m[\mu]  &  \mathcal{I}m[\nu]  & \mathcal{R}e[\mu]  & - \mathcal{R}e[\nu] \\
 \mathcal{I}m[\nu]  & \mathcal{I}m[\mu]  & - \mathcal{R}e[\nu] & \mathcal{R}e[\mu]  \\
\end{array} 
\right)
\left( 
\begin{array}{cccccccc}
\hat{x}_{s0} \\  
\hat{x}_{i0}   \\
\hat{y}_{s0} \\ 
\hat{y}_{i0}  \\
\end{array} 
\right) \nn \\
&\equiv& S_{tot} \left( 
\begin{array}{cccccccc}
\hat{x}_{s0} \\  
\hat{x}_{i0}   \\
\hat{y}_{s0} \\ 
\hat{y}_{i0}  \\
\end{array} 
\right).
\ea
Applying $S_{tot}$ to the initial covariance matrix $\Sigma_{si,0} = \mathcal{I}/4$ we obtain
\ba
\label{eq:fluttuazioni}
&&\Sigma_{si}  =  S_{tot} \Sigma_{si,0} S_{tot}^T = S_{tot}  S_{tot}^T /4 \\ 
&&  \hspace{-0.5cm} =  \fr{1}{4} \left( \begin{array}{cccccccc}
|\mu|^2 + |\nu|^2 & 2 \mathcal{R}e[\mu \nu]  & 0 & 2 \mathcal{I}m[\mu \nu]    \\  
2 \mathcal{R}e[\mu \nu]  & |\mu|^2 + |\nu|^2 &  2 \mathcal{I}m[\mu \nu]  & 0 & \\ 
0 & 2 \mathcal{I}m[\mu \nu]  & |\mu|^2 + |\nu|^2 & - 2 \mathcal{R}e[\mu \nu]   \\
2 \mathcal{I}m[\mu \nu]  & 0  & - 2 \mathcal{R}e[\mu \nu] & |\mu|^2 + |\nu|^2  \\
\end{array} 
\right)\ , \nn
\ea
which displays isotropic fluctuations for the modes $s$ and $i$, as shown in Figs. \ref{Fig05} and \ref{Fig09}. Note that these fluctuations appear to be amplified by $(G_{\mathrm{max}}+G_{\mathrm{min}})/2$, which corresponds to the classical gain  defined in Eq. (\ref{eq:gps2}) with $\theta_{\mu}-\theta_{\nu}+\theta_{s0}+\theta_{i0}= \pi/2$. This is due to the fact that the signal and idler fields are along the bisectors of the `+' and `-' fields and that the fluctuations ellipses of the `+' and `-' fields are orthogonal, as can be seen in Fig.\ \ref{Fig05}.

From Eq. (\ref{eq:fluttuazioni}) we see that the signal and idler are correlated at the output of the amplifier, as evidenced by the non zero non-diagonal elements in $\Sigma_{si}$. Due to the rotation of the squeezing direction induced by the propagation in the non-linear medium, it is instructive to consider the  \emph{rotated} sum and difference fields expressed by Eq.(\ref{eq:bmdem2p}) and (\ref{eq:bmdem2m}) in order to better understand these correlations. In that basis, as implicitly expressed in Eq. (\ref{eq:b-mvrb}), the propagator associated to the non-linear medium is diagonal, and the output covariance matrix is given by
\ba
\label{eq:fluctuations2}
 \hspace{-0.25cm} {\Sigma_{\pm}^{''}}  =  \fr{1}{4} \left( 
\begin{array}{cccccccc}
(|\mu| +  |\nu|)^2 & 0 & 0 & 0 \\
0  & (|\mu| -  |\nu|)^2 & 0 & 0 \\
0 &  0 &  (|\mu| -  |\nu|)^2 & 0 \\ 
0 & 0 & 0 &  (|\mu| +  |\nu|)^2 \\ 
\end{array} 
\right). \nn
\ea
Eq.(\ref{eq:fluctuations2}) is the diagonal version of Eq.(\ref{eq:fluttuazioni}), and can be obtained from it by using the change of basis defined in Eqs.(\ref{eq:changing1}),(\ref{eq:changing2}) when these are expressed in the quadrature basis. It represents the very well known two-mode squeezed state, with inverse squeezing degree on the two modes \cite{Furusawa_book}.

\begin{figure} [h!]
\centering
\includegraphics[width=\columnwidth]{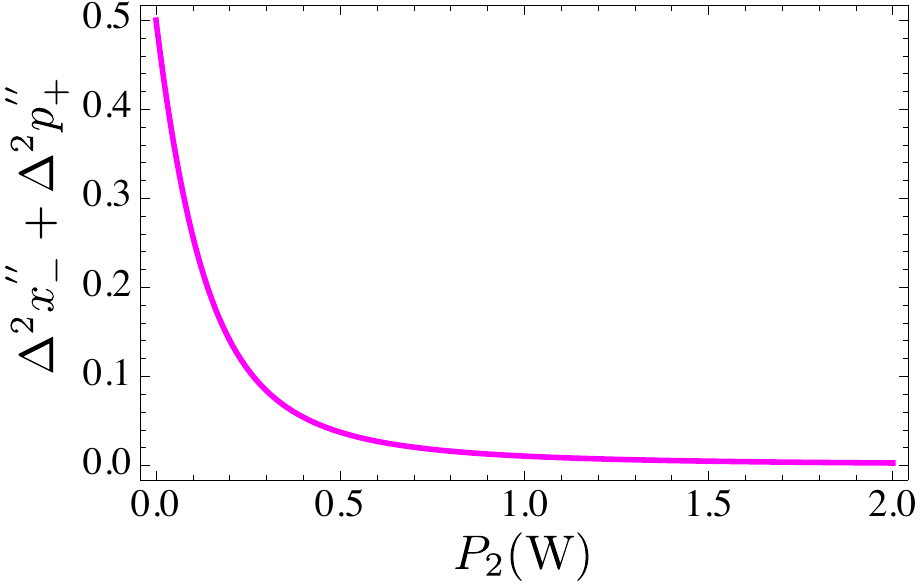}
\caption{Left hand side of the Duan criterion in Eq. (\ref{eq:Duan}), given by the sum of the second and the third diagonal terms of the matrix given in Eq. (\ref{eq:fluctuations2}) as a function of the pump power for parameters $\beta = 4.53\times10^{-11}\ \mathrm{m}^{-1}$, $z = 300\ \mathrm{m}$ , and $ \gamma = 11.3\times10^{-3}\ \mathrm{W}^{-1}. \mathrm{m}^{-1}$.}
\label{fig:correlazioni}
\end{figure}

It is hence natural that signal and idler in the rotated basis result in correlated fields, as can be inferred with the help of the Duan criterion \cite{Duan}.
This criterion implies that for two normalized modes with annihilation operators $\hat{a}_1,\hat{a}_2$ (with our convention choice $ |[\hat{x}_k,\hat{p}_k] | = 1/2$, $k=1,2$)
\be
\Delta^2 \lt \fr{\hat{x}_1 - \hat{x}_i}{\sqrt{2}} \rt + \Delta^2 \lt \fr{\hat{p}_s + \hat{p}_i}{\sqrt{2}} \rt  < \fr{1}{2}
\ee
is a sufficient condition for inseparability with respect to the bipartition $1$-$2$ (note that our modes $\hat{a}_+$ and $\hat a_-$ defined in Eqs.\,(\ref{eq:changing1},\ref{eq:changing2}) are also normalized). Hence 
\be
\label{eq:Duan}
\Delta^2 x^{''}_{-} + \Delta^2 p^{''}_{+} < \fr{1}{2}
\ee 
implies inseparability of the signal and idler modes in the rotated basis. Fig. \ref{fig:correlazioni} reproduces the evolution of the left-hand side of Eq.(\ref{eq:Duan}) versus pump power, given by the sum $\Sigma_{\pm}^{''} (2,2) + \Sigma_{\pm}^{''} (3,3)$ of the second and the third diagonal terms in the matrix of Eq. (\ref{eq:fluctuations2}). As soon as the pump power is above zero, these two modes are inseparable;  the stronger the pump, the stronger the correlations.

\section{Loss Management} \label{losses}

Experimental implementation of such an amplifier based on four-wave mixing in a nonlinear medium unavoidably leads to the apparition of losses due to propagation, fiber splicing, filters, etc. Thus, we now consider how the noise figure calculated in the previous chapter is affected by the presence of losses. We do this in two different cases: the one in which the losses occur after the amplification - which models the  propagation in a long fiber following the amplification occurring in a non-linear fiber - and the one in which the losses occur before the amplification - i.e., the non-linear fiber follows the propagation in the standard transmission line, as sketched in Fig. \ref{losses_sketch}.  We perform the calculation in a fully quantum fashion. We consider the noise figure defined by a homodyne detection. In particular, we wish to address the question whether it is more advantageous to have the amplification followed by the lossy transmission or conversely to put the amplifier after the lossy transmission line. 

\subsection{Configuration ``A''}
\label{lossesA}

\begin{figure} [h!]
\centering
\includegraphics[width=\columnwidth]{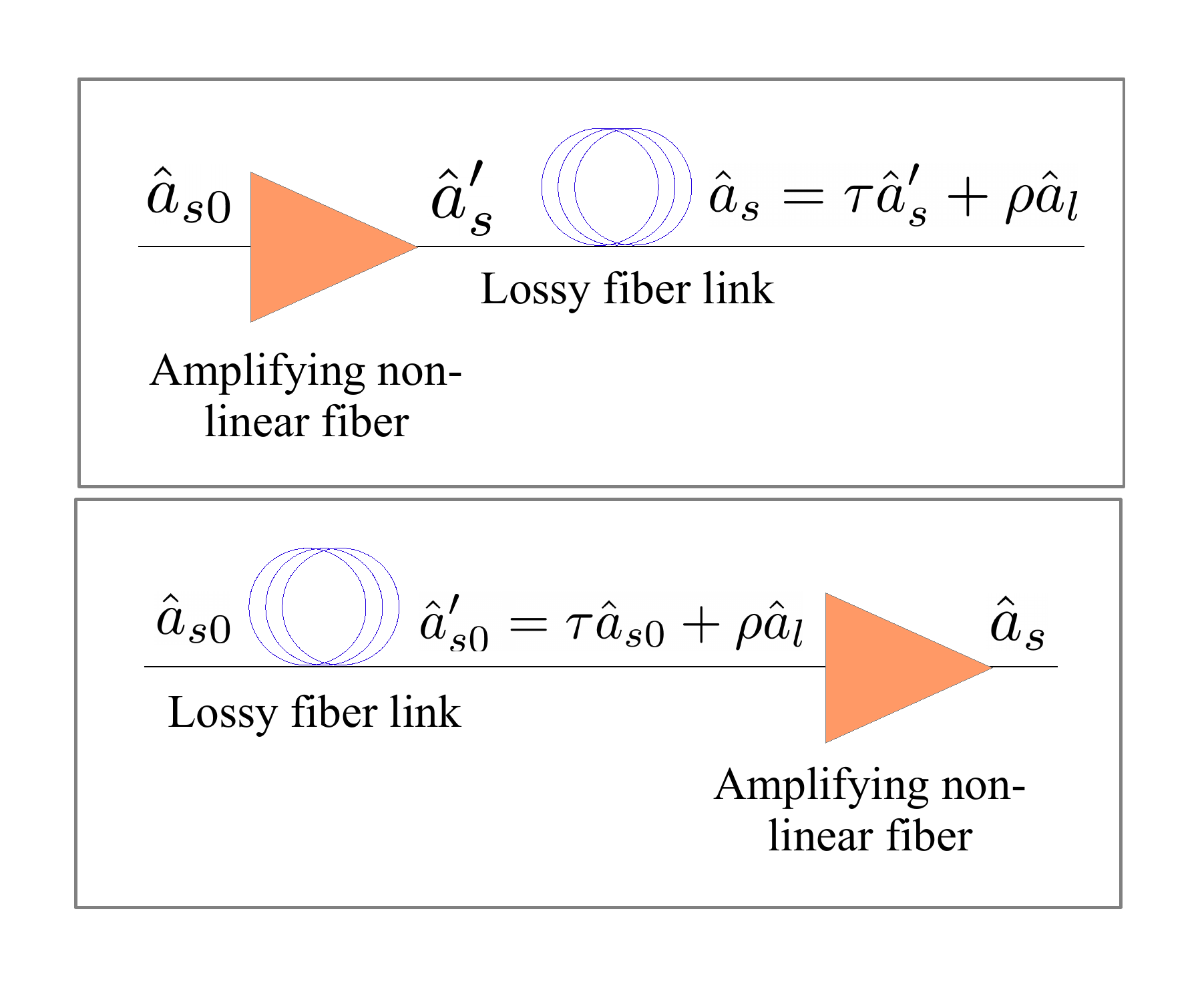}
\caption{Schematic representation of two possible links, respectively composed either by an amplifier followed by a lossy transmission link (top), or conversely by a lossy link followed by an amplifier (bottom).}
\label{losses_sketch}
\end{figure}

\ssse{Amplifier followed by losses}

The sketch of the situation we want to describe is presented in Fig. \ref{losses_sketch}, top panel.
We model the losses as the partial coupling to the extra mode ${\hat a}_{l}$, according to 
\ba
\label{eq:amp_losses1st}
{\hat a}_s &=& \tau \mu {\hat a}_{s0} + \tau \nu {\hat a}_{s0}^{\dagger} + \rho {\hat a}_{l}\ ,
\ea
where we have used Eq. (\ref{eq:solB1}) to relate the fields $\hat{a}_{s0}$ and $\hat{a}'_{s}$ in Fig.\ref{losses_sketch}, and where we have taken real values for $\tau$ and $\rho$ with $\tau^2 + \rho^2 = 1$. The same approach is carried out in Ref. \cite{McKinstrie_10}.
The input SNR assuming $|\psi_{0} \rangle = |\alpha_{s0} \rangle_s  |0 \rangle_{l}$, where we have explicited the index of the signal and loss modes in the ket expression, is given by Eq.(\ref{eq:input_snr}).
Let us evaluate the output SNR. We have from Eq.(\ref{eq:amp_losses1st})
\ba
\label{eq:exp_valueA1}
\langle  \hat x_{s \varphi}  \rangle &=& 
\fr{1}{2} \langle  \lqu   e^{-i \varphi}  \lt \tau \mu {\hat a}_{s0} + \tau \nu {\hat a}_{s0}^{\dagger} + \rho_s {\hat a}_{l} \rt +  \text{h.c.}   \rqu \rangle \nn \\
&=& \tau |\alpha_{s0} |  \lqu | \mu |  \cos (\theta_\mu + \theta_{s0} - \varphi)\right. \nn 
\\ &&+\left. |\nu |   \cos (\theta_\nu - \theta_{s0} - \varphi) \rqu  \ .
\ea
The variance is also easily computed and leads to (see Appendix \ref{app_NF})
\ba
\label{eq:var_loss_A1}
& & \Delta^2 {x_{s \varphi}} =  \\
& & \fr{\tau^2 \lqu 2 |\mu | |\nu | \cos (\theta_\mu + \theta_\nu - 2 \varphi) + (|\mu |^2 + |\nu |^2 - 1)\rqu + 1}{4}. \nn
\ea
From this we obtain the output signal-to-noise ratio as
\ba
& & \mathrm{SNR}_{\text{out}}^{\text{AL}} = \fr{\langle \hat x_{s \varphi} \rangle^2}{\Delta^2 {x_{s \varphi}}} =  \\
& &  \fr{4 \tau^2 \lgr  |\alpha_{s0} | \lqu | \mu |\cos (\theta_\mu + \theta_{s0} - \varphi) + | \nu | \cos (\theta_\nu - \theta_{s0} - \varphi) \rqu \rgr^2}{2 \tau^2 |\mu | |\nu | \cos (\theta_\mu + \theta_\nu - 2 \varphi) + 2 \tau^2 |\nu |^2 + 1},  \nn
\ea
where ``AL" stands for ``amplifier - loss" to indicate that the losses are put after the amplification.
Using  Eq. (\ref{eq:input_snr}) for the input signal-to-noise ratio this leads to the noise figure
\ba
\label{eq:NF_rewritten1st}
\hspace{-0.5cm} \mathrm{NF}^{\text{AL}}  = \fr{\mathrm{SNR}_{\text{in}}}{\mathrm{SNR}^{\text{AL}} _{\text{out}}} = 
\mathrm{NF} \lt \fr{\lambda_\varphi \tau^2 - \tau^2 + 1}{\lambda_\varphi \tau^2} \rt 
\ea
where $\mathrm{NF}$ is the lossless noise figure in the absence of losses found in Eq.\,(\ref{eq:NFnolossA}) and where we have defined
\ba
\label{eq:parametri_vari}
\lambda_\varphi &\equiv& (2 |\mu | |\nu | \cos (\theta_\mu + \theta_\nu - 2 \varphi)  + |\mu |^2 + |\nu |^2 ). \nn 
\ea

By direct derivation of Eq. (\ref{eq:NF_rewritten1st}) with respect to the detection angle $\varphi$ and the input signal phase $\theta_{s0}$ (see Appendix \ref{se:derivation}) we can see that the point of minimum of $\mathrm{NF}^{\text{AL}} $ is identified by the same coordinates as that for $\mathrm{NF}$, namely as seen in Sec. \ref{se:snrA}
\ba
\label{cond_angles}
\theta_{s0} &=& \phi = -\fr{(\theta_\mu - \theta_\nu)}{2}\ , \nn \\
\varphi &=& \theta = \fr{\theta_\mu + \theta_\nu}{2}\ ,
\ea
as expected, since in the situation depicted in Fig.\,\ref{losses_sketch} the losses do not induce changes in the mode phase.
Under this choice of the detection angle we obtain
\ba
& & \lambda_{\text{opt}} =  (| \mu | + | \nu |)^2 = G_{\text{max}}\ , \nn \\
& & \mathrm{NF}_{\text{opt}} = 1\ ,
\ea
where $G_{\text{max}}$ is the maximal classical gain of Eq. (\ref{eq:clax_gain_maxmin}).
Replacing in Eq. (\ref{eq:NF_rewritten1st}) we obtain 
\be
\label{eq:nalopt}
\mathrm{NF}^{\text{AL}}_{\text{opt}} = \lt 1 -  \fr{1}{G_{\text{max}}} + \fr{1}{G_{\text{max}} \tau^2}\rt.
\ee
This quantity will be compared to optimum noise figure in the case of a lossy transmission link followed by an amplifier in Sec.\ref{ssse-which-is-the-bestA}.
Incidentally, we notice that Eq.(\ref{eq:nalopt}) formally renders Eq.(7.45) of Ref.\,\cite{Kolobov_99} which was derived for a $\chi^{(2)}$ medium, in which the expression of the gain is different with respect to our case. There, though, a $\tau^2$ factor in the denominator is appearing since losses are in that case due to an imperfect detection, therefore equally affecting the input and the output signal-to-noise ratios.

\ssse{Losses followed by amplifier}

We model again the losses as the partial transfer of photons to the extra mode ${\hat a}_{l}$, according to equation
\ba
\label{eq:losses_amp1st}
{\hat a}_s = \tau \mu {\hat a}_{s0} + \tau \nu {\hat a}_{s0}^{\dagger} +  \mu \rho {\hat a}_{l} +  \nu \rho {\hat a}_{l}^{\dagger},
\ea
where Eq.(\ref{eq:solB1}) is used to link the fields $\hat{a}'_{s0}$ and $\hat{a}_s$ in Fig. \ref{losses_sketch} (bottom panel) and where $\tau^2 + \rho^2 = 1$. 
From Eq. (\ref{eq:losses_amp1st}) we have
\ba
\label{eq:exp_valueA2}
& & \hspace{-0.5cm} \langle  \hat x_{s \varphi}  \rangle =  \fr{1}{2} \langle  \lqu   e^{-i \varphi}  \lt \tau \mu {\hat a}_{s0} +\tau \nu {\hat a}_{s0}^{\dagger} +  \mu \rho {\hat a}_{l} +  \nu \rho {\hat a}_{l}^{\dagger} \rt +  \text{h.c.}   \rqu \rangle \nn \\
& & \hspace{-0.5cm} = \tau | \mu |  |\alpha_{s0} | \cos (\theta_\mu + \theta_{s0} - \varphi) +\tau | \nu |  |\alpha_{i0} | \cos (\theta_\nu - \theta_{s0} - \varphi).  \nn \\
\ea
In Appendix \ref{app_NF} we compute the variance, yielding
\be
\label{eq:var_loss_A2}
\Delta^2 {x_{s \varphi}} = \fr{ | \mu | | \nu | \cos{(\theta_\mu + \theta_\nu -  2 \varphi)}}{2} + \fr{ |\nu |^2}{2} + \fr{1}{4}.
\ee
From this we obtain the output SNR as
\ba
& & \mathrm{SNR}_{\text{out}}^{\text{LA}} = \fr{\langle \hat x_{s \varphi} \rangle^2}{\Delta^2 {x_{s \varphi}}} =  \\
& &  \fr{4  \tau^2 |\alpha_{s0} |^2  \lgr | \mu |  \cos (\theta_\mu + \theta_{s0} - \varphi) + | \nu |  \cos (\theta_\nu - \theta_{s0} - \varphi) \rgr^2}{2  | \mu | | \nu | \cos{(\theta_\mu + \theta_\nu -  2 \varphi)}+ |\mu |^2 + |\nu |^2}, \nn 
\ea
where ``LA" stands for ``loss - amplifier" to indicate that the amplification is put after the lossy transmission.
Using Eq. (\ref{eq:input_snr}) for the input SNR  leads to the noise figure
\ba
\label{F21st}
\mathrm{NF}^{\text{LA}} = \fr{\mathrm{SNR}_{\text{in}}}{\mathrm{SNR}^{\text{LA}}_{\text{out}}} = \fr{\mathrm{NF}}{\tau^2}.
\ea

From Eq.(\ref{F21st}) we see that at each constant $\tau$ the noise figure in the presence of losses is optimal when the one in the absence of losses is, namely for the input conditions Eq. (\ref{cond_angles}), for which $\mathrm{NF} = 1$.  Hence
\be
\label{eq:nlaopt}
\mathrm{NF}^{\text{LA}}_{\text{opt}} = \fr{1}{\tau^2}.
\ee

\ssse{Which is the best choice?}
\label{ssse-which-is-the-bestA}

Let us consider now the ratio between the two optimal noise figures in the case of a link composed of an amplifier and a lossy transmission or a lossy transmission followed by an amplifier, expressed respectively by Eqs. (\ref{eq:nalopt}) and (\ref{eq:nlaopt}). This yields
\be
\label{eq:ratio_nalopt}
\fr{\mathrm{NF}^{\text{AL}}_{\text{opt}}}{\mathrm{NF}^{\text{LA}}_{\text{opt}} }  =  \fr{G_{\text{max}} \tau^2 - \tau^2 + 1}{G_{\text{max}}}.
\ee
Due to the fact that $\tau^2 \leq 1$, we see that as soon as $G_{\text{max}} > 1$ the ratio in Eq.\,(\ref{eq:ratio_nalopt}) is less than one, i.e. the configuration with the amplifier followed by the lossy link is more convenient.
This result seem natural since putting the amplifier after the losses amplifies the vacuum fluctuations of the loss mode as well. This is the same conclusion as the one usually obtained for a phase insensitive amplifier \cite{Desurvire2002}.  

\subsection{Configuration ``B''}
\label{lossesB}

  We have seen in Sec.\ref{se:snrB} that in the large gain limit the noise figure in configuration ``B"  can evidence noiseless amplification even by considering signal-to-noise ratios with respect to the signal only, despite a non-zero input idler power, due to the output signal-idler correlations. Hence in this section we are going to consider noise figures with respect to the signal only for simplicity. The presence of losses does not change this argument when these occur before the amplification (Sec.\ref{lossesB2}) but may degrade the signal-idler correlation if they occur after amplification, as in Sec.\ref{lossesB1}. We though assume that this effect is negligible in the large gain limit.

\ssse{Amplifier followed by losses}
\label{lossesB1}

An analogous situation as presented in Fig. \ref{losses_sketch}, top panel, can be considered for configuration ``B", with now two input and output modes.
In this case as well we can model the losses as the partial transfer of photons to extra modes, noted ${\hat a}_{l1}$ and ${\hat a}_{l2}$, according to 
\ba
\label{eq:amp_losses}
{\hat a}_s &=& \tau_s \mu {\hat a}_{s0} + \tau_s \nu {\hat a}_{i0}^{\dagger} + \rho_s {\hat a}_{l1}\ , \nn \\
{\hat a}_i &=& \tau_i \nu {\hat a}_{s0}^{\dagger} + \tau_i \mu {\hat a}_{i0} + \rho_i {\hat a}_{l2}\ ,
\ea
where we have used Eq. (\ref{solA1}) to express the amplification and where $\tau_j$ and $\rho_j$ are real with $\tau_j^2 + \rho_j^2 = 1$ for $j = s,i$.
The input SNR has already been computed in Eq. (\ref{eq:input_snr}) assuming $|\psi_{0} \rangle = |\alpha_{s0} \rangle_s |\alpha_{i0} \rangle_i |0 \rangle_{l1} |0 \rangle_{l2}$. Let us evaluate the output SNR. We have from Eq. (\ref{eq:amp_losses})
\ba
\label{eq:exp_value}
  \hspace{-0.35cm}\langle  \hat x_{s \varphi}  \rangle & =& \fr{1}{2} \langle  \lqu   e^{-i \varphi}  \lt \tau_s \mu {\hat a}_{s0} + \tau_s \nu {\hat a}_{i0}^{\dagger} + \rho_s {\hat a}_{l1} \rt +  \text{h.c.}   \rqu \rangle \nn  \\
 & =& \tau_s \lqu | \mu |  |\alpha_{s0} | \cos (\theta_\mu + \theta_{s0} - \varphi)\right.\nn\\
 && \left. + | \nu |  |\alpha_{i0} | \cos (\theta_\nu - \theta_{i0} - \varphi) \rqu \ .
\ea
The calculation of the variance performed in Appendix \ref{app_NF} gives
\be
\label{eq:var_B1}
\Delta^2 {x_{s \varphi}} = \langle  \hat x_{s \varphi}^2  \rangle - \langle  \hat x_{s \varphi}  \rangle^2 = \fr{\tau_s^2 |\nu |^2}{2} + \fr{1}{4},
\ee
which results isotropic (i.e., independent on $\varphi$) as in the lossless case, consistently with the fact that in the simple loss model of Fig.\,\ref{losses_sketch} the coupling to the loss mode does not introduce dephasing.
From Eqs. (\ref{eq:exp_value}) and (\ref{eq:var_B1}) we obtain the output SNR as
\ba
\hspace{-0.2cm} & & \mathrm{SNR}^{\text{LA}}_{\text{out}} = \fr{\langle \hat x_{s \varphi} \rangle^2}{\Delta^2 {x_{s \varphi}}} =  \fr{4}{(2 \tau_s^2 |\nu |^2 + 1)} \times \\
\hspace{-0.25cm} & &  \hspace{-0.3cm} \lgr \tau_s \lqu | \mu |  |\alpha_{s0} | \cos (\theta_\mu + \theta_{s0} - \varphi) + | \nu |  |\alpha_{i0} | \cos (\theta_\nu - \theta_{i0} - \varphi) \rqu \rgr^2.\nn
\ea
Using the expression in Eq. (\ref{eq:input_snr}) for the input SNR and the fact that $2  |\nu |^2 +1 = |\nu |^2 + |\mu |^2$ leads to the noise figure
\ba
\label{eq:NF_rewritten}
\mathrm{NF}^{\text{LA}} = \mathrm{NF} \lt \fr{G_0 \tau_s^2 - \tau_s^2 + 1}{G_0 \tau_s^2} \rt\ ,
\ea
where NF is the noise figure in the absence of losses provided in Eq.\,(\ref{eq:NFnolossB}) and  where we have introduced 
\ba
\label{eq:defs_par}
G_0 = (|\mu |^2 + |\nu |^2 ).
\ea
Note that the latter parameter corresponds to the classical gain defined in Eq.\,(\ref{eq:gps2}), with $\theta_{\mu}-\theta_{\nu}+\theta_{s0}+\theta_{i0}= \pi/2$.
Also note that Eq. (\ref{eq:NF_rewritten}) is formally equivalent to Eq. (\ref{eq:NF_rewritten1st}), i.e. to the noise figure that we had found in the configuration with two pumps, apart from the different definitions of $NF$ and of the gain parameter $G_0$. 

It is easily seen that in Eq. (\ref{eq:NF_rewritten}) $\mathrm{NF}^{\text{LA}}$ depends on $\varphi$ and $\theta_{s0}$ only because $\mathrm{NF}$ does. Hence, $\mathrm{NF}^{\text{LA}}$ is optimized when $\mathrm{NF}$ is optimized, and this happens when 
\ba
\label{optcond100}
\theta_{i0} &=& - \theta_{s0} - (\theta_\mu - \theta_\nu)\ ,\nn \\
\varphi &=& \theta_\mu + \theta_{s0},
\ea
as we have seen in Sec.\ref{se:snrB}, yielding $\mathrm{NF} = 1$ after correction by a two factor to take into account the input idler power ($P_i = P_s$).
Replacing Eq. (\ref{optcond100}) in Eq. (\ref{eq:NF_rewritten}) gives again
\be
\label{eq:ajdkbduuyg}
\mathrm{NF}^{\text{LA}}  = \lt 1 -  \fr{1}{G_0} + \fr{1}{G_0 \tau_s^2} \rt .
\ee
This result is analogous to what we had found for the amplifier - loss transmission link in the configuration with two pumps (see Eq. (\ref{eq:nalopt})).
This should be compared to Eq. (111) of Ref. \cite{McKinstrie_10}. In that case, however, the noise figure is computed for a direct detection and for $|\alpha_{i0} |^2 = 0$. 

\ssse{Losses followed by amplifier}
\label{lossesB2}

We model again the losses as the partial transfer of photons to the extra modes ${\hat a}_{l1}$ and ${\hat a}_{l2}$, according to 
\ba
\label{eq:losses_amp}
{\hat a}_s &=& \tau_s \mu {\hat a}_{s0} + \tau_i \nu {\hat a}_{i0}^{\dagger} +  \mu \rho_s {\hat a}_{l1} +  \nu \rho_i {\hat a}_{l2}^{\dagger}\ , \nn \\
{\hat a}_i &=& \tau_s \nu {\hat a}_{s0}^{\dagger} + \tau_i \mu {\hat a}_{i0} + \nu \rho_s {\hat a}_{l1}^{\dagger} + \mu \rho_i {\hat a}_{l2}\ ,
\ea
where $\tau_j$ and $\rho_j$ are real with $\tau_j^2 + \rho_j^2 = 1$ for $j = s,i$. 
From Eq. (\ref{eq:amp_losses}) we have
\ba
\label{eq:exp_value2}
   \langle  \hat x_{s \varphi}  \rangle &= & \fr{1}{2} \langle  \lqu   e^{-i \varphi}  \lt \tau_s \mu {\hat a}_{s0} + \tau_i \nu {\hat a}_{i0}^{\dagger} \right. \right.\nn \nn\\
 && \left. \left.+  \mu \rho_s {\hat a}_{l1} +  \nu \rho_i {\hat a}_{l2}^{\dagger} \rt +  \text{h.c.}   \rqu \rangle  \nn \\
 &=& \tau_s | \mu |  |\alpha_{s0} | \cos (\theta_\mu + \theta_{s0} - \varphi) \nn \\
 &&+ \tau_i | \nu |  |\alpha_{i0} | \cos (\theta_\nu - \theta_{i0} - \varphi)  \ .
\ea
The variance reads (see Appendix \ref{app_NF})
\be
\label{eq:var_B2}
\Delta^2 {x_{s \varphi}} =  \fr{ |\nu |^2}{2} + \fr{1}{4} = \fr{|\mu |^2 + |\nu |^2}{4}.
\ee
From this we obtain the output SNR as
\ba
  \mathrm{SNR}_{\text{out}} &=& \fr{\langle \hat x_{s \varphi} \rangle^2}{\Delta^2 {x_{s \varphi}}}  \nn \\
&= &   \fr{1}{(|\mu |^2 + |\nu |^2)} \left\{4 \left[ \tau_s | \mu |  |\alpha_{s0} | \cos (\theta_\mu + \theta_{s0} - \varphi)\right.\right.\nn\\
&&\left. \left.+ \tau_i | \nu |  |\alpha_{i0} | \cos (\theta_\nu - \theta_{i0} - \varphi) \right]^2 \right\} \ ,
\ea
which, using Eq. (\ref{eq:input_snr}) for the input SNR,  leads to the noise figure
\be
\label{eq:F2rewritten}
\hspace{-0.1cm} \mathrm{NF}^{\text{LA}}  = \fr{\mathrm{NF}}{\tau^2} 
\ee 
where we have assumed the physically meaningful condition $\tau_s = \tau_i \equiv \tau$.
Note that Eq. (\ref{eq:F2rewritten}) is formally identical to Eq. (\ref{eq:nlaopt}) that we had found for the corresponding case in the first configuration.
From Eq. (\ref{eq:F2rewritten}) we immediately see that at each constant loss parameter $\tau$, the noise figure in the presence of losses is optimal when the noise figure in the absence of losses is. This happens for the input conditions expressed by Eq. (\ref{optcond100}), yielding again one (when taking into account the idler input power as well).
Hence we have
\be
\label{eq:nlaoptbis}\ 
\mathrm{NF}^{\text{LA}}_{\text{opt}} = \fr{1}{\tau^2} ,
\ee
similarly to Eq. (\ref{eq:nlaopt}). 
This should be compared to Eq. (117) of Ref. \cite{McKinstrie_10}. In that case, however, the noise figure is computed for a direct detection and for $\alpha_{i0} = \alpha_{s0}$. 

\ssse{Which is the best choice?}
 
Since the two expressions for $\mathrm{NF}^{\text{LA}}_{\text{opt}}$ and $\mathrm{NF}^{\text{AL}}_{\text{opt}}$ are both formally analogous to the corresponding expressions in the configuration with a single pump (with $G_{\text{max}}$ replaced by $G_0$), then the ratio of the two quantities exactly yields the  expression in Eq. (\ref{eq:ratio_nalopt}), namely
\be
\label{eq:ratio_naloptbis}
\fr{\mathrm{NF}^{\text{AL}}_{\text{opt}}}{\mathrm{NF}^{\text{LA}}_{\text{opt}} }  =  \fr{G_0 \tau^2 - \tau^2 + 1}{G_0}.
\ee
Analogously to the preceding case, as soon as $G_0 > 1$ the ratio in Eq. (\ref{eq:ratio_naloptbis}) goes below one, i.e. the configuration with the amplifier followed by the lossy link is more efficient.
As we have already notices, this appears natural since putting the amplifier after the loss mode amplifies the vacuum fluctuations of the loss mode as well. 

\section{Conclusions} 

We have shown that the symplectic formalism is useful to describe the transformation governing the fields evolution in a parametric noiseless amplifier. Indeed this has allowed us to interpret the fields  evolution in terms of a squeezing operation on the relevant modes (the signal mode in configuration A and the symmetric and antisymmetric combinations of signal and idler in configuration B), preceded and followed by a quadrature rotation which depends on the input phases as well as on the fiber parameters (among others, its length). Incidentally, this provides a natural explanation to understand the rotation of the noise ellipse generated via four wave mixing reported in Ref. \cite{Lett_Glorieux _13}. Furthermore, this has allowed us to characterize the correlations between signal and idler in a simple way - as off-diagonal elements of the covariance matrix in the signal-idler basis, as well as in terms of the Duan criterion.
These correlations have revealed possible to characterize the amplifier noise figure in the large gain limit by measuring only the signal  at the output, despite a non zero idler output power.
 We have then analyzed the noise figure of the amplifier in both configurations, in the absence and in the presence of losses, showing that it is always preferable to make the amplification preceding the lossy transmission. 

\section*{Acknowledgements}
We thank C. Fabre for useful discussions.
This work was partially supported by the Agence Nationale de la Recherche (Project NAMOCS No. ANR-12-BS03-001-01),  Thales Research \& Technology, and Thales Airborne Systems. 

\appendix

\section{Solution of the four-wave mixing equation} 
\subsection{Configuration A}
\label{app:solA}

In the undepleted pump approximation where $|A_{1,3}|^2 \equiv P_{1,3}$ is constant Eq.(\ref{eq:4-wave-mixing} a-c) 
have solution $A_1(z) = A_1(0) e^{i \gamma (P_1 + 2 P_3)}, A_3(z) = A_3(0) e^{i \gamma (2 P_1 +  P_3)} $.
Substituting in Eq.(\ref{eq:4-wave-mixing} b) yields
\be
\label{eq:diff_eq_rewrA}
\fr{d A_2}{dz} = i 2 \gamma \lqu (P_1 +  P_3) A_2 +   A_1(0) A_3(0) e^{i 3 \gamma (P_1 + P_3)} A_2^* e^{-i \beta z} \rqu.
\ee
Introducing the field $B_2 = A_2 e^{-i 2 \gamma  (P_1 +  P_3) z} $ Eq.(\ref{eq:diff_eq_rewrA}) becomes
\be
\label{eq:field_B0bis_derbis}
\fr{d B_2}{dz} = i 2 \gamma A_1(0) A_3(0)  e^{-i \kappa z} B_2^*
\ee
where we have used that 
$A_2^* = B_2^* e^{-i 2 \gamma  (P_1 +  P_3) z} $, and where we have introduced $\kappa = (\beta + \gamma (P_1 + P_3))$.
Differentiating a second  time Eq.(\ref{eq:field_B0bis_derbis}) leads to
\be
\fr{d^2 B_2}{d^2z} = i 2 \gamma A_1(0) A_3(0)  \lt  \fr{d B_2^*}{dz} - i \kappa B_2^* \rt e^{-i \kappa z} .
\ee
We now use that $\fr{d B_2^*}{dz} = - 2i \gamma A_1^*(0) A_3^*(0)  e^{i \kappa z} B_2$
and that from Eq.(\ref{eq:field_B0bis_derbis}) $i 2 \gamma A_1(0) A_3(0)  (-i  \kappa) B_2^* e^{-i \kappa z} =   (-i  \kappa) \fr{d B_2}{dz}$, obtaining
\be
\label{eq:diff_sec_conf}
\fr{d^2 B_2}{d^2z}  + i  \kappa \fr{d B_2}{dz} -  4  \gamma^2 P_1 P_3 B_2 = 0.
\ee
The solution of Eq.(\ref{eq:diff_sec_conf}) is
\be
\label{eq:solution_2c}
B_2(z) =  \lt e^{- g z} a +  e^{ g z} b  \rt e^{\fr{- i \kappa z}{2}}
\ee
with $g = \sqrt{4 \gamma^2 P_1 P_3 - (\fr{\kappa}{2})^2 }$.
We now pose the initial conditions. Using the definition of $B_2$
\ba
\label{eq:initial_condA}
& \hspace{-0.5cm}   B_2(0) = A_2(0) = a + c; \\
 & \hspace{-0.5cm} \lt \fr{d B_2}{dz} \rt_{z =0} \hspace{-0.15cm}  = 
 i 2 \gamma A_1(0) A_3 (0) A_2^*(0)  =  -a \lt  g + i\fr{\kappa}{2} \rt + c \lt g - i\fr{\kappa}{2} \rt.  \nn
\ea
From Eq.(\ref{eq:initial_condA}) we find
\ba
\label{eq:initial_condbis_2c}
a = \fr{1}{2} A_1(0) \lt 1 -  i\fr{\kappa}{2g} \rt - \fr{ i \gamma A_1 (0) A_3 (0) A_2^*(0)}{g}; \nn \\
c = \fr{1}{2} A_1(0) \lt 1 +  i\fr{\kappa}{2g} \rt + \fr{ i \gamma A_1 (0) A_3 (0) A_2^*(0)}{g}.
\ea
Substituting Eq.(\ref{eq:initial_condbis_2c}) in (\ref{eq:solution_2c})
and re-expressing everything in terms of the fields $A_i$ only, also using the definition of the parameter $\kappa$, finally leads to Eq.(\ref{eq:solB1}) of the main text.
 
\subsection{Configuration B}
\label{app:solB}

In the undepleted pump approximation $|A_2|^2 \equiv P_2$ constant 
has solution $A_2(z) = A_2(0) e^{i \gamma P_2 z}$.
Substituting in Eqs.(\ref{eq:4-wave-mixing}-a,c) yields
\ba
\label{steps_app}
\fr{d A_1}{dz} &= i \gamma \lt 2 P_2  A_1 + A_2^2(0) e^{ 2i \gamma P_2 z} A_3^* e^{i \beta z} \rt  & a) \nn \\
\fr{d A_3}{dz} &= i \gamma \lt  2 P_2 A_3 +  A_2^2(0) e^{ 2i \gamma P_2 z}  A_1^*   e^{i \beta z} \rt. & b).
\ea
Introducing the fields $B_1 = A_1 e^{-i 2 \gamma P_2 z}$ and $B_3 = A_3 e^{-i 2 \gamma P_2 z}$ and substituting in Eq.(\ref{steps_app})
yields
to the set of coupled equations
\ba
\label{eq:field_B}
\fr{d B_1}{dz} &=  i \gamma A_2^2(0) B_3^* e^{- i \kappa z} & a)\nn \\
\fr{d B_3}{dz} &=  i \gamma A_2^2(0) B_1^* e^{- i \kappa z}  & b)
\ea
where we have introduced the parameter $\kappa =  2 \gamma P_2 - \beta$. Deriving a second time the first line of Eq.(\ref{eq:field_B}) gives
\be 
\label{eq:steps} 
\fr{d^2 B_1}{dz^2} =  i \gamma A_2^2(0)  \lt \fr{d B_3^*}{dz} - i \kappa B_3^*  \rt  e^{-i \kappa z}. 
\ee
We now use that 
$\fr{d B_3^*}{dz} =  -i \gamma A_2^{2*}(0) B_1 e^{i \kappa z}$
and that from Eq.(\ref{eq:field_B}-a) we have $- i^2 \gamma  \kappa A_2^2(0) B_3^* e^{-i \kappa z} = - i  \kappa \fr{d B_1}{dz}$, obtaining from Eq.(\ref{eq:steps})
\be
\label{eq:eq-diff}
\fr{d^2 B_1}{dz^2} +  i  \kappa \fr{d B_1}{dz} -   \gamma^2 P_2^2 B_1 = 0.
\ee
The solution of Eq.(\ref{eq:eq-diff}) is
\be
\label{eq:solution}
B_1(z) = \lt e^{- g z} a +  e^{ g z} b  \rt e^{-\fr{i \kappa z}{2}}
\ee 
with $g = \sqrt{ \gamma^2 P_2^2 - (\fr{\kappa}{2})^2 }$.
We now pose the initial conditions. Using the definition of $B_1$ and $B_3$
\ba
\label{eq:initial_cond}
& \hspace{-0.4cm}  B_1(0) = A_1(0) = a + c; \\
& \hspace{-0.4cm}  \lt \fr{d B_1}{dz} \rt_{z =0} 
=  i \gamma A_2^2 (0) A_3^*(0)  = - a \lt g +  i\fr{\kappa}{2} \rt + c \lt g - i\fr{\kappa}{2} \rt. \nn 
\ea
From Eq.(\ref{eq:initial_cond}) we find
\ba
\label{eq:initial_condbis}
a = \fr{1}{2} A_1(0) \lt 1 -  i\fr{\kappa}{2g} \rt - \fr{ i \gamma A_2^2 (0) A_3^*(0)}{2 g}; \nn \\
c = \fr{1}{2} A_1(0) \lt 1 +  i\fr{\kappa}{2g} \rt + \fr{ i \gamma A_2^2 (0) A_3^*(0)}{2 g}.
\ea
Substituting Eq.(\ref{eq:initial_condbis}) in (\ref{eq:solution}) yields
and re-expressing all in terms of the fields $A_i$ only, using newly the definition of the parameter $\kappa$, finally leads to Eq.(\ref{solA1}) of the main text.

\se{Derivation of the coefficients of the Bloch-Messiah decomposition from the experimental parameters}\label{AppendixD}

\subsection{Configuration ``A''}
By inserting Eqs.\ (\ref{eq:solB2}) and (\ref{eq:solB3}) into Eqs.\ (\ref{eq:sol_angles3}) and (\ref{eq:sol_angles4}), we obtain the following expressions for the parameters of the Bloch-Messiah decomposition:
\ba
\tan(\theta_{\mu}+\theta_{\nu})&=&\frac{\frac{\kappa}{2g}\tanh gz \tan \theta_g-1}{\frac{\kappa}{2g}\tanh gz + \tan \theta_g}=\tan 2\theta\ ,\\
\tan(\theta_{\mu}-\theta_{\nu})&=&\frac{\frac{\kappa}{2g}\tanh gz \tan (\theta_{10}+\theta_{30})+1}{-\frac{\kappa}{2g}\tanh gz + \tan (\theta_{10}+\theta_{30})}=-\tan 2\phi\ , \nn\\&&
\ea
where
\be
\theta_g=\theta_{10}+\theta_{30}+2\delta z\ .
\ee
\subsection{Configuration ``B''}
In the case of configuration ``B'' with one degenerate pump, inserting Eqs.\ (\ref{solA2}) and (\ref{solA3}) into Eqs.\ (\ref{eq:sol_angles3}) and (\ref{eq:sol_angles4}) leads to:
\ba
\label{eq:micro_expr}
\tan(\theta_{\mu}+\theta_{\nu})&=&\frac{\frac{\kappa}{2g}\tanh gz \tan \theta_g-1}{\frac{\kappa}{2g}\tanh gz + \tan \theta_g}=\tan 2\theta\ ,\\
\tan(\theta_{\mu}-\theta_{\nu})&=&\frac{\frac{\kappa}{2g}\tanh gz \tan 2\theta_{20}+1}{-\frac{\kappa}{2g}\tanh gz + \tan 2\theta_{20}}=-\tan 2\phi\ , \nn\\&&
\ea
where
\be
\theta_g=2\theta_{20}+\Delta\beta z\ .
\ee
\se{Calculation of the noise figure in the presence and absence of losses }\label{app_NF}

We proceed with the calculation of the variances given in Sec.\ref{se:snrA},\ref{se:snrB} and \ref{losses} of the main text.
The variance is defined according to the expression
\be
\Delta^2 {x_{s \varphi}} = \langle  \hat x_{s \varphi}^2  \rangle -  \langle  \hat x_{s \varphi}  \rangle^2.
\ee
Variances are easily computed using that
\ba
\label{quadr_op1st}
\langle  \hat x_{s \varphi}^2  \rangle = \fr{1}{4} \langleÊ( \hat a_s^2 e^{-i 2 \varphi} + \hat  {\hat a}_{s}^{\dagger} e^{i 2 \varphi} + 2 \hat  {\hat a}_{s}^{\dagger} \hat a_s + 1) \rangle,
\ea
the expectation values $\langle  \hat x_{s \varphi}  \rangle$ being given in the main text.

\sse{Configuration ``A" in the absence of losses}

Let us start with the variance computed in Sec.\ref{se:snrA} , i.e. for configuration ``A" and in the absence of losses.
From Eq.(\ref{eq:amp_losses1st}) and Eq.(\ref{quadr_op1st}) we obtain 
\ba
\label{eq:steps_int_var1st}
\hspace{-0.1cm} \langle  \hat a_s^2  \rangle &=&   \mu^2 \alpha_{s0}^2 + \nu^2 \alpha_{s0}^{*2}  + 2 \mu \nu \alpha_{s0} \alpha_{i0}^{*} +  \mu \nu  \\
\hspace{-0.1cm} \langle  \hat a_s^{\dagger 2}  \rangle &=& \mbox{\, \, (complex conjugate)} \nn \\
\hspace{-0.1cm} \langle  \hat a_s^{\dagger} \hat a_s  \rangle &=&  |\mu |^2 |\alpha_{s0}  |^2 +   |\nu |^2 (|\alpha_{s0}  |^2 + 1) + 2 \mathcal{R}e [\mu^* \nu \alpha_{s0}^{*2} ].  \nn
\ea
Substituting Eq.(\ref{eq:steps_int_var1st}) in Eq.(\ref{quadr_op1st})  we obtain in few steps 
\ba
\label{square_quadr1st}
\langle  \hat x_{s \varphi}^2  \rangle  &=& |\alpha_{s0} |^2 \lgr | \mu |  |\alpha_{s0} | \cos (\theta_\mu + \theta_{s0} - \varphi) \right. \nn \\
&+& \left. | \nu |   \cos (\theta_\nu - \theta_{s0} - \varphi) \rgr^2  \\
&+& \fr{ |\mu | |\nu | \cos (\theta_\mu + \theta_\nu - 2 \varphi)}{2} +\fr{ |\nu |^2}{2} + \fr{1}{4}. \nn
\ea
Subtraction of the expectation value $\langle  \hat x_{s \varphi}  \rangle^2$ in Eq.(\ref{eq:823}) leads to Eq.(\ref{Eq78}) of the main text.

\sse{Configuration ``B" in the absence of losses}
\label{app:fluctuations_mode_output} 

We compute here the fluctuations of the mode defined in Eq.(\ref{eq:output_amp9}).
The transformation which brings from the signal and idler modes to the mode sum defined in Eq.(\ref{eq:output_amp9}) (and an auxiliary mode difference which will not be relevant for our discussion) is given by a phase space  rotation and by performing the sum (and difference) of the rotated modes. Explicitly,
\ba
\label{eq:trasf_modes88}
\left( 
\begin{array}{cccccccc}
\hat{x}_{s \theta_s} \\  
\hat{x}_{i \theta_i}   \\
\hat{y}_{s \theta_s} \\ 
\hat{y}_{i \theta_i}  \\
\end{array} 
\right) \hspace{-0.13cm} &=& \hspace{-0.13cm}
\left( 
\begin{array}{cccccccc}
\cos \theta_s & 0 & \sin \theta_s & 0 \\  
0 & \cos \theta_i & 0 & \sin \theta_i  \\ 
-\sin \theta_s & 0 & \cos \theta_s & 0 \\  
0 & -\sin \theta_i & 0 & \cos \theta_i  \\ 
\end{array} 
\right)
\left( 
\begin{array}{cccccccc}
\hat{x}_{s} \\  
\hat{x}_{i}   \\
\hat{y}_{s} \\ 
\hat{y}_{i}  \\
\end{array} 
\right); \nn \\
\left( 
\begin{array}{cccccccc}
\hat{x} \\  
\hat{x}_{\text{diff}}   \\
\hat{y} \\ 
\hat{y}_{\text{diff}}   \\
\end{array} 
\right) \hspace{-0.13cm} &=& \hspace{-0.13cm} \fr{1}{\sqrt{2}}
\left( 
\begin{array}{cccccccc}
1 & 1 & 0 & 0 \\  
1 & -1 & 0 & 0  \\ 
0 & 0 & 1 & 1 \\  
0 & 0 & 1 & -1 \\ 
\end{array} 
\right)
\left( 
\begin{array}{cccccccc}
\hat{x}_{s \theta_s} \\  
\hat{x}_{i \theta_i}   \\
\hat{y}_{s \theta_s} \\ 
\hat{y}_{i \theta_i}  \\
\end{array} 
\right)  \nn
\ea
which combined gives 
\ba
\label{eq:trasf_modes77}
\hspace{-0.13cm}  \left( 
\begin{array}{cccccccc}
\hat{x}_{s \theta_s} \\  
\hat{x}_{i \theta_i}   \\
\hat{y}_{s \theta_s} \\ 
\hat{y}_{i \theta_i}  \\
\end{array} 
\right) \hspace{-0.13cm} &=& \hspace{-0.13cm} \fr{1}{\sqrt{2}}
\left(
\begin{array}{cccc}
\cos \theta_s &\cos \theta_i &\sin \theta_s & \sin \theta_i \\
\cos \theta_s & -\cos \theta_i &\sin \theta_s & - \sin \theta_i \\
 - \sin \theta_s & - \sin \theta_i &\cos \theta_s & \cos \theta_i \\
 - \sin \theta_s &\sin \theta_i &\cos \theta_s & - \cos \theta_i
\end{array}
\right)
\left( 
\begin{array}{cccccccc}
\hat{x}_{s} \\  
\hat{x}_{i}   \\
\hat{y}_{s} \\ 
\hat{y}_{i}  \\
\end{array} 
\right) \nn \\
&\equiv& S_{\text{transf}}  \left( 
\begin{array}{cccccccc}
\hat{x}_{s} \\  
\hat{x}_{i}   \\
\hat{y}_{s} \\ 
\hat{y}_{i}  \\
\end{array} 
\right).
\ea
The application of the transformation $S_{\text{transf}}$ to the output covariance matrix in the signal-idler basis given in Eq.(\ref{eq:fluttuazioni}) brings to the diagonal covariance matrix $S_{\text{transf}}  \Sigma_{si} S_{\text{transf}}^T$, where the first element represents the fluctuations of interest, resulting in Eq.(\ref{eq:flutt_mode_interest}) of the main text.

\sse{Configuration ``A"}

We turn now to the calculation of the variances in the presence of losses, either put before of after the amplification, in each of the two possible configurations.
\ssse{Amplifier followed by losses}

We proceed with the calculation of the variance in Eq.(\ref{eq:parametri_vari}) of the main text.
The explicit calculation is easy since we can use Eqs.(\ref{eq:amp_losses1st}) and (\ref{quadr_op1st}), where all the terms containing ${\hat a}_{l}^2$ or ${\hat a}_{l}^{\dagger 2}$ or  ${\hat a}_{l}^{\dagger} {\hat a}_{l}$ give zero contribution. Hence we obtain 
\ba
\label{eq:steps_int_var1st}
\langle  \hat a_s^2  \rangle &=&  \tau^2 \mu^2 \alpha_{s0}^2 +  \tau^2 \nu^2 \alpha_{s0}^{*2}  + 2 \tau^2 \mu \nu \alpha_{s0} \alpha_{i0}^{*} +  \tau^2 \mu \nu \nn \\
\langle  \hat a_s^{\dagger 2}  \rangle &=& \mbox{\, \, (complex conjugate)} \\
\langle  \hat a_s^{\dagger} \hat a_s  \rangle &=& \tau^2   |\mu |^2 |\alpha_{s0}  |^2 + \tau^2  |\nu |^2 (|\alpha_{s0}  |^2 + 1) \nn \\
& & + 2 \tau^2 \mathcal{R}e [\mu^* \nu \alpha_{s0}^{*2} ].  \nn
\ea
Substituting Eq.(\ref{eq:steps_int_var1st}) in Eq.(\ref{quadr_op1st})  we obtain in few steps 
\ba
\label{square_quadr1st}
\langle  \hat x_{s \varphi}^2  \rangle  &=& \tau^2  |\alpha_{s0} |^2 \lgr | \mu | \cos (\theta_\mu + \theta_{s0} - \varphi) \right. \nn \\
&+& \left. | \nu | \cos (\theta_\nu - \theta_{s0} - \varphi) \rgr^2  \\
&+& \fr{\tau^2 |\mu | |\nu | \cos (\theta_\mu + \theta_\nu - 2 \varphi)}{2} +\fr{\tau^2 |\nu |^2}{2} + \fr{1}{4}. \nn
\ea
It is easy to see by that from Eqs.(\ref{eq:exp_valueA1}) and (\ref{square_quadr1st}) and using that $2  |\nu |^2 +1 = |\nu |^2 + |\mu |^2$ we obtain Eq.(\ref{eq:var_loss_A1}) of the main text.

\ssse{Losses followed by amplifier}

We now derive Eq.(\ref{eq:var_loss_A2}) of the main text. We use Eqs.(\ref{eq:losses_amp1st}) and (\ref{quadr_op1st}) to compute the variance. All the terms involving the loss mode give zero contribution when projected on the vacuum loss input state, except $\mu \nu \rho^2 \hat{a}_l \hat{a}_l^{\dagger}$ in $\hat a_s^2$ and $| \nu |^2 \rho_i^2 \hat{a}_{l}\hat{a}_{l}^{\dagger}$  in $a_s^{\dagger} \hat a_s$. We obtain hence
\ba
\label{eq:steps_int_var2}
 \langle  \hat a_s^2  \rangle  &=& \tau^2 \mu^2 \alpha_{s0}^2 +\tau^2 \nu^2 \alpha_{s0}^{*2}  + 2 \mu \nu \tau^2  | \alpha_{s0}  |^2 +  \mu \nu \tau^2 + \mu \nu  \rho^2 \nn \\  
\langle  \hat a_s^{\dagger 2}  \rangle &=&  \mbox{\, \, (complex conjugate)}    \\
\langle  \hat a_s^{\dagger} \hat a_s  \rangle &=& \tau^2 | \mu |^2  | \alpha_{s0} |^2 + \tau^2  | \nu |^2 ( | \alpha_{s0} |^2 + 1) \nn \\
& & + 2\tau^2  \mathcal{R}e [\mu^* \nu  \alpha_{s0}^{*2}]  + | \nu |^2 \rho^2.    \nn  
\ea
Substituting Eq.(\ref{eq:steps_int_var2}) in Eq.(\ref{quadr_op1st}) we obtain in some step
\ba
\label{square_quadr2}
\langle  \hat x_{s \varphi}^2  \rangle \hspace{-0.1cm} &=& \hspace{-0.1cm} \tau^2 |\alpha_{s0} |^2 \lgr  | \mu |  \cos (\theta_\mu + \theta_{s0} - \varphi) \right. \nn \\
& & \left. + | \nu |  \cos (\theta_\nu - \theta_{s0} - \varphi)\rgr^2   \\ 
& & + \fr{| \mu | | \nu | \cos{(\theta_\mu + \theta_\nu -  2 \varphi)}}{2} + \fr{ |\nu |^2}{2} + \fr{1}{4}.  \nn
\ea
It is easy to see by Eqs.(\ref{eq:exp_valueA2}) and (\ref{square_quadr2}) that we obtain Eq.(\ref{eq:var_loss_A2}) of the main text.

\sse{Configuration ``B"}

\ssse{Amplifier followed by losses}

We proceed with the calculation of Eqs.(\ref{eq:var_B1}) of the main text.
Using Eq.(\ref{eq:amp_losses})  and (\ref{quadr_op1st}), the calculation is easy since all the terms containing ${\hat a}_{l1}^2$ or ${\hat a}_{l1}^{\dagger 2}$ or  ${\hat a}_{l1}^{\dagger} {\hat a}_{l1}$ give zero contribution. Hence we obtain 
\ba
\label{eq:steps_int_var}
\langle  \hat a_s^2  \rangle &=&  \tau_s^2 \mu^2 \alpha_{s0}^2 +  \tau_s^2 \nu^2 \alpha_{i0}^{*2}  + 2 \tau_s^2 \mu \nu \alpha_{s0} \alpha_{i0}^{*} \nn \\
\langle  \hat a_s^{\dagger 2}  \rangle &=& \mbox{\, \, (complex conjugate)} \\
\langle  \hat a_s^{\dagger} \hat a_s  \rangle &=& \tau_s^2  |\mu |^2 |\alpha_{s0}  |^2 +  \tau_s^2  |\nu |^2 (|\alpha_{i0}  |^2 + 1) \nn \\
& & + 2 \tau_s^2 \mathcal{R}e [\mu^* \nu \alpha_{s0}^*  \alpha_{i0}^* ]. \nn
\ea
Substituting Eq.(\ref{eq:steps_int_var}) in Eq.(\ref{quadr_op1st}) we obtain in few steps
\ba
\label{square_quadr}
\langle  \hat x_{s \varphi}^2  \rangle &=& \tau_s^2 \lgr | \mu |  |\alpha_{s0} | \cos (\theta_\mu + \theta_{s0} - \varphi) \right.  \\
& & \left. + | \nu |  |\alpha_{i0} | \cos (\theta_\nu - \theta_{i0} - \varphi) \rgr^2 + \fr{\tau_s^2 |\nu |^2}{2} + \fr{1}{4}, \nn
\ea
It is easy to see by Eqs.(\ref{eq:exp_value}) and (\ref{square_quadr}) that one obtain Eq.(\ref{eq:var_B1}).

\ssse{Losses followed by amplifier}

We now use Eqs. (\ref{eq:losses_amp}) and (\ref{quadr_op1st}) to compute the variance in Eq.(\ref{eq:var_B2}). All the terms involving the loss modes give zero contribution when projected on the vacua loss input state, except the one $| \nu |^2 \rho_i^2 \hat{a}_{l2}\hat{a}_{l2}^{\dagger}$  in $a_s^{\dagger} \hat a_s$. We obtain hence
\ba
\label{eq:steps_int_var2}
\langle  \hat a_s^2  \rangle &=& \tau_s^2 \mu^2 \alpha_{s0}^2 +\tau_i^2 \nu^2 \alpha_{i0}^{*2}  + 2 \mu \nu \tau_s \tau_i  \alpha_{s0} \alpha_{i0}^*  \nn \\  
\langle  \hat a_s^{\dagger 2}  \rangle &=&  \mbox{\, \, (complex conjugate)}   \\
\langle  \hat a_s^{\dagger} \hat a_s  \rangle &=& \tau_s^2  | \mu |^2  | \alpha_{s0} |^2 + \tau_i^2  | \nu |^2 ( | \alpha_{i0} |^2 + 1) \nn \\
& & + 2 \mathcal{R}e [\mu^* \nu \tau_s \tau_i  \alpha_{s0}^* \alpha_{i0}^*]  + | \nu |^2 \rho_i^2.    \nn  
\ea
Note that in the equation in the last line we can rewrite the term $\tau_i^2  | \nu |^2 | \alpha_{i0} |^2 + \tau_i^2  | \nu |^2 + | \nu |^2 \rho_i^2 = | \nu |^2 (\tau_i^2 | \alpha_{i0} |^2 + 1)$.
Substituting Eq.(\ref{eq:steps_int_var2}) in Eq.(\ref{quadr_op1st}) we obtain in some step
\ba
\label{square_quadr2}
\langle  \hat x_{s \varphi}^2  \rangle &=& \lgr \tau_s | \mu |  |\alpha_{s0} | \cos (\theta_\mu + \theta_{s0} - \varphi) \right. \\ 
& & \left. + \tau_i | \nu |  |\alpha_{i0} | \cos (\theta_\nu - \theta_{i0} - \varphi)\rgr^2 + \fr{ |\nu |^2}{2} + \fr{1}{4}.  \nn
\ea
It is easy to see by Eqs.(\ref{eq:exp_value2}) and (\ref{square_quadr2}) that we re-obtain Eq.(\ref{eq:var_B2}).

\se{Optimization of the noise figure for configuration ``A", amplifier followed by losses} \label{se:derivation}

We now ask the question compute which initial condition on the input signal phase $\theta_{s0}$ and which detection phase $\varphi$ are optimizing the noise figure given in Eq.(\ref{eq:nalopt}) in the presence of losses.
We simply have to set
\ba
\lgr
\begin{array}{cccc}
1. \hspace{0.2cm} \fr{d}{d \theta_{s0}}NF^{\text{AL}} (\theta_{s0}, \varphi) &=& 0 \nn \\
2. \hspace{0.2cm} \fr{d}{d \varphi}NF^{\text{AL}} (\theta_{s0}, \varphi) &=& 0
\end{array} \right.
\ea
We have from Eq.(\ref{eq:NF_rewritten1st}):
{\small \be
1. \, \fr{d}{d \theta_{s0}}NF^{\text{AL}} (\theta_{s0}, \varphi) = \fr{d}{d \theta_{s0}} NF(\theta_{s0}, \varphi) \lqu  1 -  \fr{1}{\lambda(\varphi)} + \fr{1}{\lambda(\varphi) \tau^2} \rqu. \nn
\ee}
This is zero either
\begin{itemize}
\item for $\fr{d}{d \theta_{s0}} NF(\theta_{s0}, \varphi) = 0 \hspace{0.2cm} \\
\Rightarrow \hspace{0.2cm} (\theta_{s0} = -\fr{(\theta_\mu - \theta_\nu)}{2}, \varphi =  \fr{(\theta_\mu + \theta_\nu)}{2})$ (see Sec.\ref{se:snrA}).
\item for $\lqu  1 -  \fr{1}{\lambda_\varphi} + \fr{1}{\lambda(\varphi) \tau^2} \rqu = 0 \Rightarrow   \lambda(\varphi) =  \fr{\tau^2 - 1}{\tau^2}= -\fr{\rho^2}{\tau^2}$ No solution ($\lambda(\varphi) \geq  0$ always).
\end{itemize}
{\small  \ba
2. & & \fr{d}{d \varphi}NF^{\text{AL}} (\theta_{s0}, \varphi) =  \fr{d}{d \varphi}NF (\theta_{s0}, \varphi) \lqu  1 -  \fr{1}{\lambda(\varphi)} + \fr{1}{\lambda(\varphi) \tau^2} \rqu \nn \\
& & + NF  \lqu \fr{\lambda'(\varphi)(\tau^2 - 1)}{\lambda^2(\varphi) \tau^2} \rqu \nn
\ea}
with $\lambda'(\varphi) = d \lambda(\varphi)/d\varphi = 4 |\mu | |\nu | \sin (\theta_\mu + \theta_\nu - 2 \varphi) $. We easily see that the solution which renders zero the derivative with respect to $\theta_{s0}$ also renders zero $\lambda'(\varphi)$ and hence $\fr{d}{d \varphi}NF^{\text{AL}} (\theta_{s0}, \varphi)$ (since $\fr{d}{d \theta_{s0}} NF(\theta_{s0}, \varphi) = 0$ in that point as seen in 1.)
Hence we conclude that the minimum is found for the coordinates in Eq.(\ref{cond_angles}).



\begin{thebibliography}{33}
\bibitem{Footnote} Note that since the expression of $\phi$ in terms of the fiber parameters reported in Eq.\,(\ref{eq:micro_expr}) also involves the pump phases, the signal phase is in this way unambigously defined.
\end{thebibliography}
\end{document}